\documentclass{comnet}

\usepackage{amsmath} 

\usepackage{cite}
\usepackage{subfig} 
\usepackage{graphics} 
\usepackage{graphicx}
\usepackage{url} 
\usepackage{xcolor}
\usepackage{color}

\begin{document}

\title{Network depth: identifying median and contours in complex networks.}

\shorttitle{Network Depth}
\shortauthorlist{G. Bertagnolli, C. Agostinelli and M. De Domenico}

\author{
\name{Giulia Bertagnolli$^*$}
\address{Center for Information and Communication Technology, Fondazione Bruno Kessler, Via Sommarive 18, 38123 Povo (TN), Italy}
\address{Department of Mathematics, University of Trento, Via Sommarive, 14, 38123 Povo (TN), Italy\email{$^*$Corresponding author: giulia.bertagnolli@unitn.it}}
\name{Claudio Agostinelli}
\address{Department of Mathematics, University of Trento, Via Sommarive, 14, 38123 Povo (TN), Italy}
\and
\name{Manlio De Domenico$^\dag$}
\address{Center for Information and Communication Technology, Fondazione Bruno Kessler, Via Sommarive 18, 38123 Povo (TN), Italy}\email{$^\dag$Corresponding author: mdedomenico@fbk.eu}}

\maketitle

\begin{abstract}
{Centrality descriptors are widely used to rank nodes according to specific concept(s) of importance. Despite the large number of centrality measures available nowadays, it is still poorly understood how to identify the node which can be considered as the ``centre'' of a complex network. In fact, this problem corresponds to finding the median of a complex network.
The median is a nonparametric -- or better, distribution-free -- and robust estimator of the location parameter of a probability distribution. In this work, we present the statistical and most natural generalisation of the concept of median to the realm of complex networks, discussing its advantages for defining the centre of the system and percentiles around that centre. To this aim, we introduce a new statistical data depth and we apply it to networks embedded in a geometric space induced by different metrics.
The application of our framework to empirical networks allows us to identify central nodes which are socially or biologically relevant.
}{centrality; statistical data depths; diffusion geometry}
\\
\end{abstract}

\section{Introduction}
Nodes centrality is a very fundamental concept in network science, used to answer questions like ``who are the most influential actors in a social network''? Or ``which are the key nodes for the optimal functioning of a power grid''?
Unfortunately centrality descriptors are defined upon a non-rigorous definition of \textit{importance}.
Consequently, there is a plethora of centrality measures and interpretations making difficult to understand which one could be used as a one-fits-all solution.

Similarly many statistical problems deal with the location, the centre of a population. Knowing very little about the underlying population, e.g. we may not know if the underlying distribution has finite first moment, we can still estimate the location through the median and set up distribution-free statistical tests for it. Nonparametric statistics largely rely on order statistic, which is fine in $\mathbb{R}$, but gives some problems already in $\mathbb{R}^2$, where no natural order relation is defined. As we will see in Sec.~\ref{sec:depth}, statistical data depths enable the generalisation of order and order statistics, to multivariate spaces, functional spaces and, with this work, also to networks.
Before moving to our new centrality, we have to summarise some key concepts about network centrality.

\begin{figure}[!h]
\begin{minipage}{.5\linewidth}
\centering
\subfloat[\textit{1-path}]{\includegraphics[width=.6\textwidth]{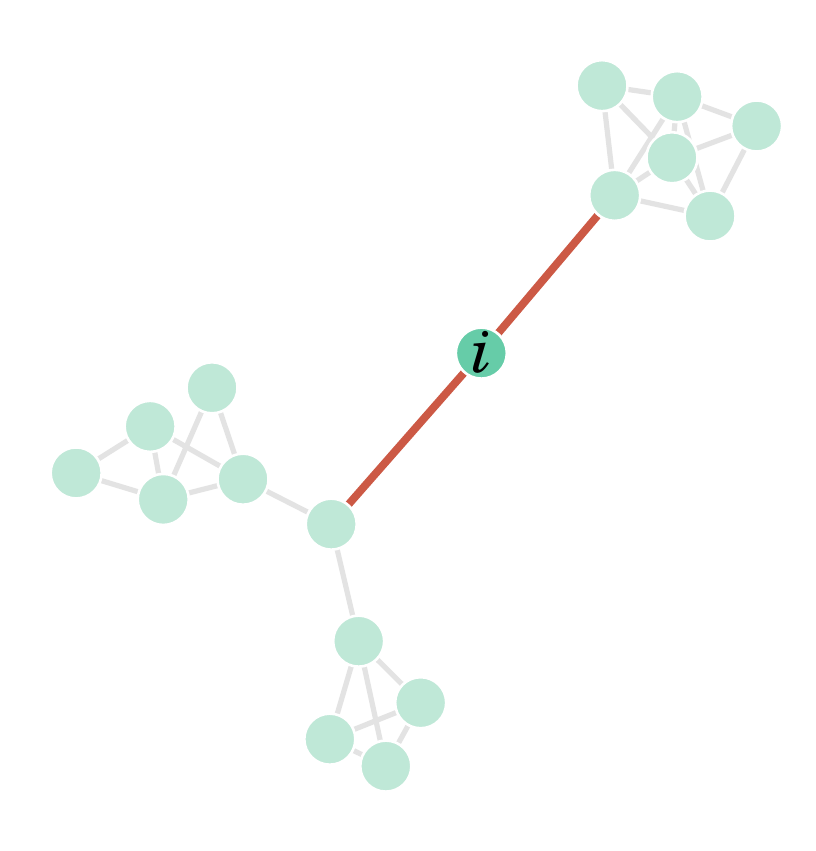}}
\end{minipage}%
\begin{minipage}{.5\linewidth}
\subfloat[\textit{Shortest-paths}]{\includegraphics[width=.6\textwidth]{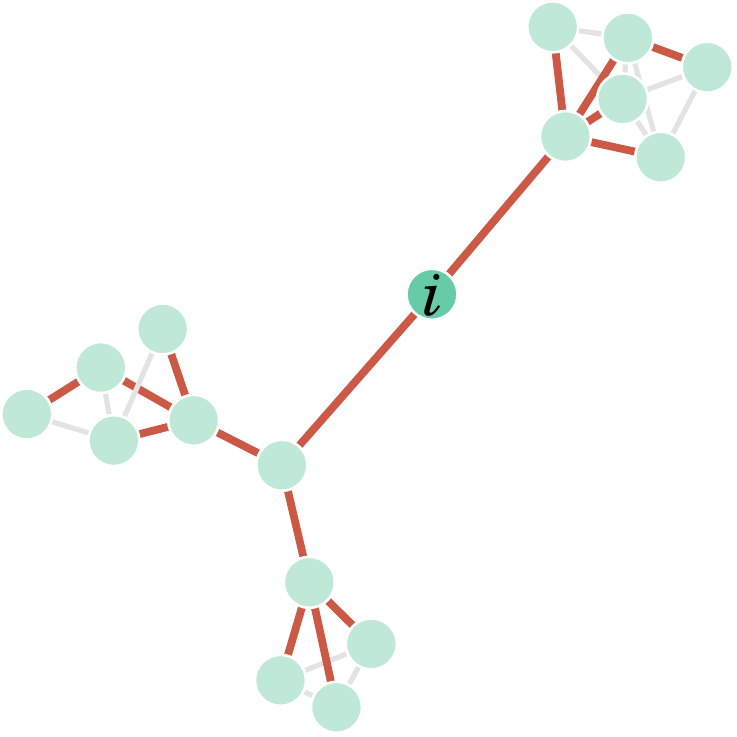}}
\end{minipage} \par\medskip
\begin{minipage}{.5\linewidth}
\centering
\subfloat[\textit{Degree}]{\label{sfig:deg}\includegraphics[width=.6\textwidth]{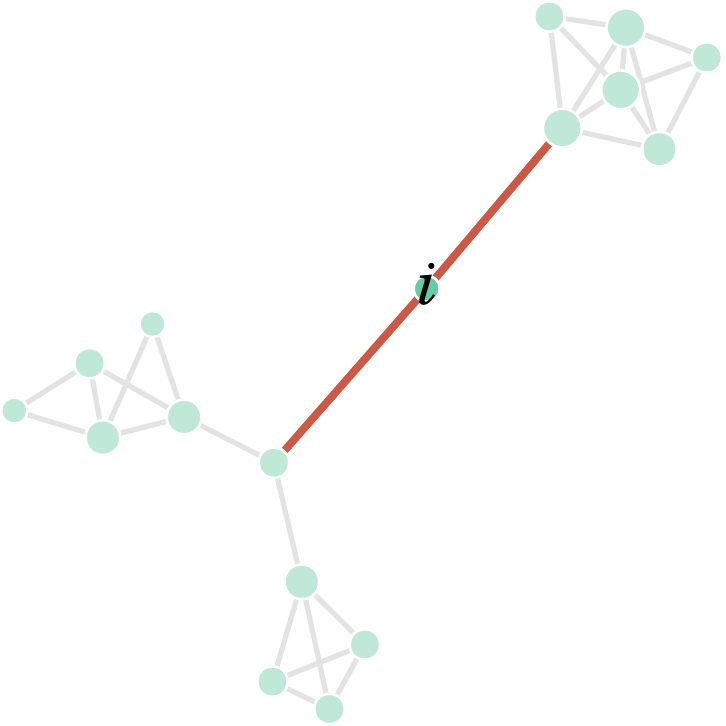}}
\end{minipage}%
\begin{minipage}{.5\linewidth}
\subfloat[\textit{Betweenness}]{\label{sfig:btw}\includegraphics[width=.6\textwidth]{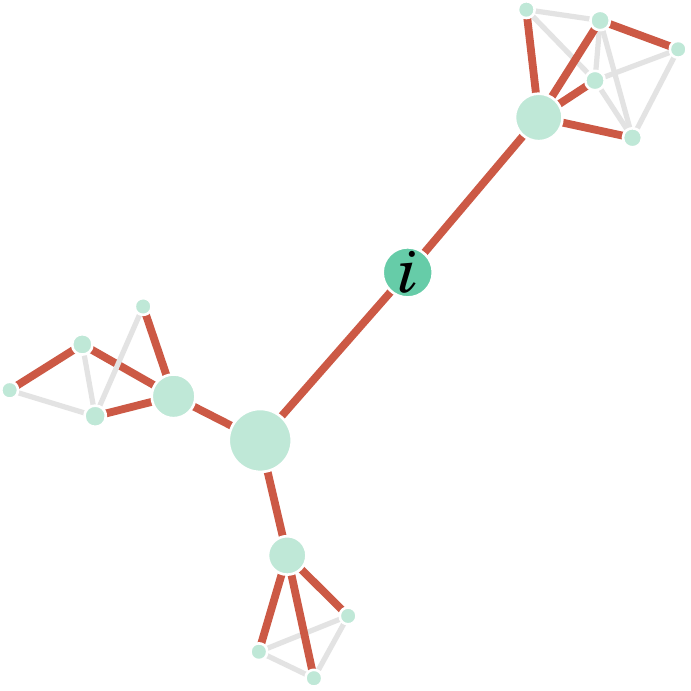}}
\end{minipage}%
\caption{Different Walk Types emanating from $i$ and two \textit{volume} centrality measures taking very diverse values for $i$. In panels (\ref{sfig:deg}) and (\ref{sfig:btw}), node size depends on the centrality score.}
\label{fig: Walk-Types}
\end{figure}

The need for a unified framework for centrality measures was already clear in the early days of network science, when Freeman proposed his three-fold categorisation of centralities \citep{Freeman1978}.
More recently, Borgatti and Everett \citep{Borgatti2006} approached centrality from a graph-theoretic perspective.
They showed that all centralities evaluate the involvement of a node in the walkability of the network according to four features: Walk Type, Walk Property, Walk Position and Summary Type.
Different Walk Types (e.g. shortest-paths or random walks) and restrictions on them (e.g. maximum length) might influence centrality as shown in Fig.~\ref{fig: Walk-Types}.
For instance the degree $k_i = \sum\limits_j a_{ij}$ of the $i$--th node can be seen as counting the paths of length 1 emanating from $i$.
Allowing paths of length $k$ leads to $k$--path centrality.
Another example is given by the betweenness that, as the degree, belongs to the class of measures that count walks and have \textit{volume} as the Walk Property, at variance with closeness that, instead, deals with paths \textit{length}.

The Walk Position can be \textit{radial}, where walks emanate from a given node -- as for degree and closeness -- or \textit{medial}, when the focus is on walks passing through that node (e.g. betweenness).
Once the information based on these features is gathered in a node-by-node matrix $W \in \R^{N \times N}$, a summary statistic $T: \R^{N \times N} \to \R^N$ is needed to obtain a centrality score vector $C = T(W)$. $T$ represents the fourth feature: Summary Type.
As observed in \citep{Borgatti2006}, the variation along this feature is small since usually the summary is a row sum $T(W) = W \textbf{1}$ or a (weighted) average.
Table~\ref{tab:cms-topology} places some popular centrality measures in this classification with respect to Walk Property and Position.
We briefly recall the definitions of the five measures we use from now on, in bold font in Tab.~\ref{tab:cms-topology}, by using the same notation. In the following, let $A = \{a_{ij}\}_{i,j=1}^N$ be the $N \times N$ adjacency matrix of graph $G = (V, E)$.

The \textit{degree} of vertex $i$ is $C_{\text{deg}}(i) = (W \textbf{1})_{i}$ where $W = A$.
Observing that in some occasions one's importance should depend on whether its neighbours are themselves important, many variations on the degree have been proposed.
In 1972 Bonacich introduced the eigenvector centrality, which can be regarded as an elegant summary of Katz’s, Hoede’s and Hubbell’s measures \citep{Borgatti2006}.
Let $\textbf{x}$ be the eigenvector corresponding to the largest eigenvalue of $A$, i.e. $A\textbf{x} = \lambda_1 \textbf{x}$; then $\textbf{x}$ is the \textit{eigenvector centrality} vector and for each $i \in 1, \dots, N$
\begin{equation}
  x_i = \lambda_1^{-1}\sum_{j=1}^N a_{ij} x_j.
\end{equation}
The entries of $W$ are then $w_{ij} = a_{ij} x_j$ and $C_{\text{eig}} = T(W) = \lambda_1^{-1} W \textbf{1}$.

The \textit{closeness} is obtained as the inverse of the marginals of $W = D_{sp}$, the shortest-path distance matrix. The elements of its centrality score vector are computed as follows
\begin{equation}
  C_{clo}(i) = \frac{1}{\sum_{j\neq i}d_{sp}(i, j)} \quad \forall i \in 1, \cdots, N.
\end{equation}
Observe that $\sum\limits_{j\neq i}d_{sp}(i, j)$ is the total sp-distance from node $i$ to all others.
Sometimes the average distance $\frac{1}{N}\sum\limits_{j\neq i}d_{sp}(i, j)$ or $\frac{1}{N-1}\sum\limits_{j\neq i}d_{sp}(i, j)$ \citep{Freeman1978} is considered to make the scores comparable among networks with different sizes.

The \textit{betweenness} centrality of node $i$, $C_{btw}(i)$, is the total number of shortest-paths between pairs of vertices that pass through node $i$.

Borgatti's \textit{Distance-weighted Fragmentation (DF)}\footnote{Available in the R package ``keyplayer'' \citep{An2016}} \citep{Borgatti2003} measures the extent to which the network cohesiveness depends on node $i \in \{1, \dots, N\}$.
It depends both on the size of components of the network after the removal of $i$ and on the relative cohesion of each component, in terms of total sp-distance between all pairs of nodes.
Let $d = \max\limits_{j,k \neq i} d_{sp}^{-1}(j, k)$
\begin{equation}
  C_{\text{DF}}(i) = 1 - \frac{\sum\limits_{i\neq j, k = 1}^N d_{sp}^{-1}(j, k)}{d N (N-1)}.
\end{equation}

\begin{table}[t]
\begin{tabular}{l|ll}
   & Radial           & Medial \\ \hline
Volume & Freeman degree, Bonacich eigenvector & Betweenness \\
Length & closeness      & Borgatti DF
\end{tabular}
\caption{Table adapted from \citep{Borgatti2006}. Walk Property and Walk Position dimensions and some popular centralities classified on the basis of this two features.}
\label{tab:cms-topology}
\end{table}

We have chosen this five measures, since they are representative of other centralities.

The remainder of this paper is organised as follows. In Section \ref{sec:depth} we present a new centrality built upon robust, nonparametric statistics, called statistical data depths.
In particular, we start introducing the statistical background of data depths and the definition of our new multivariate depth function, the Projected Tukey depth (PTD).
In Subsection \ref{ssec:network-depth} we make use of the graph-theoretical framework for centralities by Borgatti and Everett, to show that the PTD evaluated on a network embedded in space defines a new centrality in the sense of networks. We call this new centrality \textit{network depth}.
In addition, this framework allows differences and commonalities among the network depth and other centrality descriptors to be identified theoretically.
In Subsection \ref{ssec:net-emb} we examine in detail the network embedding step in our procedure and study how it reflects on the network depth. In fact, networks can be embedded in different ways. We use Multidimensional scaling (MDS), a metric preserving embedding method that depends on two parameters: a distance (or dissimilarity) matrix and an embedding dimension. This step introduces a parameter space and, consequently, a family of network depths.
We provide some applications to empirical networks in Section \ref{sec:applications}, together with the heuristics that should be followed to reproduce our results and to evaluate the network depth to real data. The conclusions in Section \ref{sec:conclusions} summarise our contribution to the understanding of the concept of centre-outward ordering of nodes in a network.

\section{Network Depth} \label{sec:depth}

A statistical data depth function, or simply a \textit{depth}, can be seen as a measure of the centrality or outlyingness of a given sample with respect to its underlying probability distribution.
In the analysis of multivariate data, it allows us to find an overall centre of the observations, which plays the role of the median.
Further, depths induce a centre-outward and probability-based ordering of sample points, which results in the ranking of the observed units.

Based on this ordering, quantitative and graphical univariate tools for data analysis have been generalised to the multivariate case \citep{Regina1999}.
The usefulness of depths is widely recognised, indeed they have been extended to functional data as well \citep{Nieto2016}.

\begin{figure}[!h]
\subfloat[1D]{\includegraphics[width=.7\textwidth]{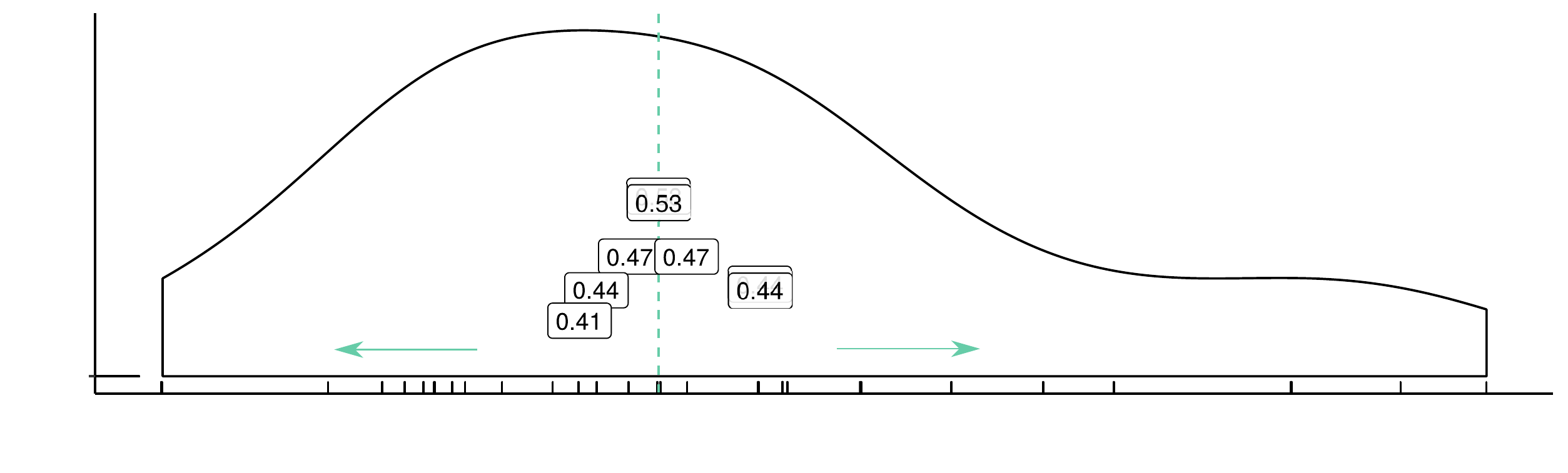}}
\subfloat[2D]{\includegraphics[width=.28\textwidth]{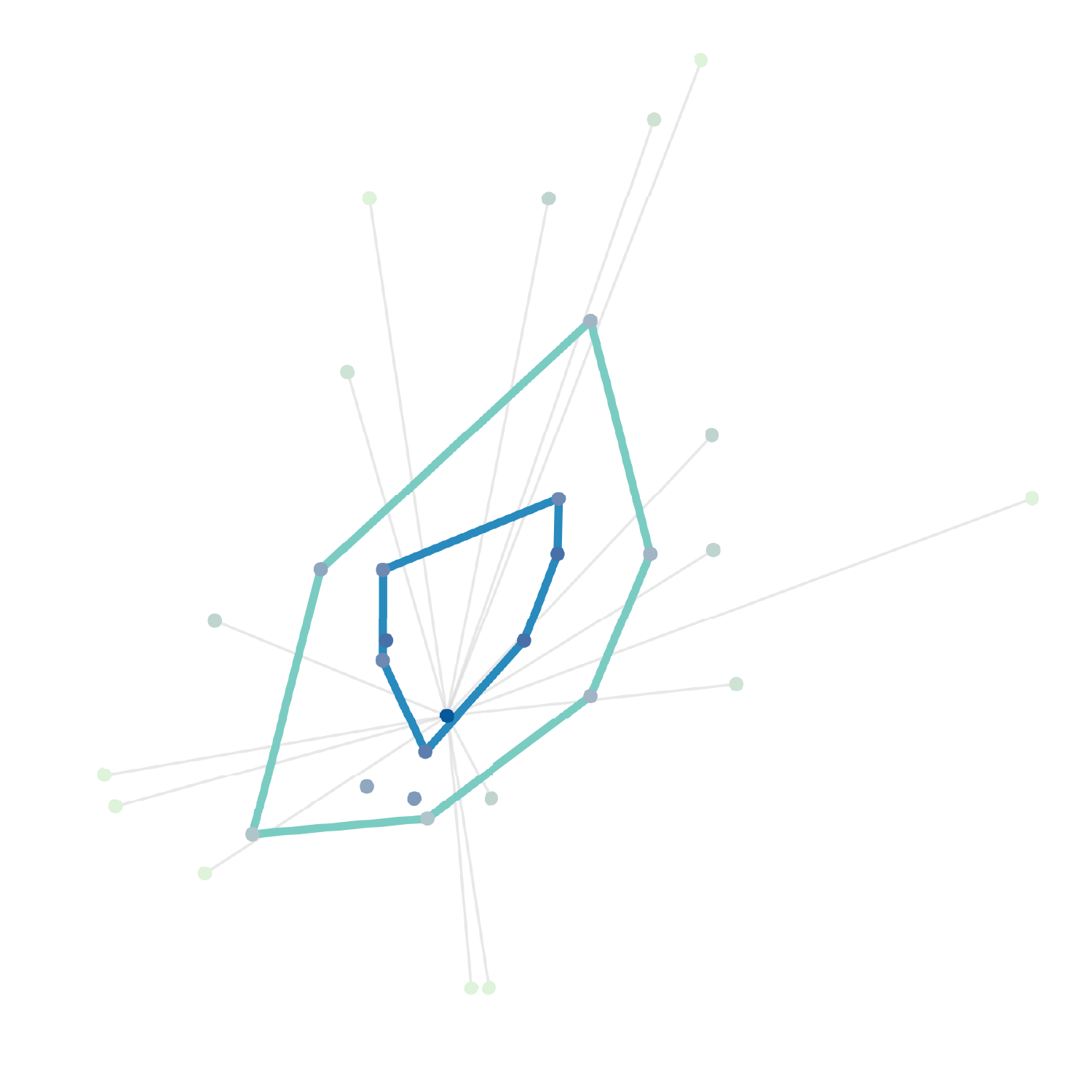}}
\caption{Centre-outward ordering of sample points. The multivariate median is the deepest point, while order statistics (quantiles) are generalised through depth-induced contours, which in the sample case can be estimated through convex hulls containing the most deep (central) $(1-\alpha)100\%$ sample points.}
\label{fig:fig-sim}
\end{figure}

The idea underling data depths dates back to 1975 when John W. Tukey \citep{Tukey1975} proposed a new approach to the problem of ordering points in $\R^p$.
Tukey's idea, which was further developed in \citep{Donoho1992}, is based on the observation that, given a univariate random variable $X$ with distribution function $F_X$, a natural centre-outward ordering of points can be derived as follows: for each $x \in \R$ we can compute $\mathbb{P}(X \leq x) = F_X(x)$ and $\mathbb{P}(X \geq x) = 1 - F_X(x)$ and define $D_1(x, F_X) = \min\{\mathbb{P}(X \leq x), \mathbb{P}(X \geq x)\}$.
The median of $X$ -- the value $m$ such that $\mathbb{P}(X \leq m), \mathbb{P}(X \geq m) \geq 0.5$ -- is the most central (or the deepest) point, while points far away from it in both directions have smaller depth values, see Fig.~\ref{fig:fig-sim}.
Then, the \textit{halfspace} Tukey depth ($\HDe$)  of a point $\textbf{x} \in \R^p$ w.r.t. the distribution of a random vector $\textbf{X}$, is the least depth value of $\textbf{x}$ in any one-dimensional projection of $\textbf{X}$.

Many depths and corresponding empirical estimators have been proposed since, but it was only in 2000 that statistical depth functions have been rigorously defined.
In \citep{Zuo2000a} Zuo and Serfling state four desirable properties a function should fulfil in order to be called a \textit{depth}.
These formalise the intuition that the concept of ``centre'' is relevant. Therefore, if a probability distribution is \textit{symmetric}\footnote{There exist different definitions of symmetric probability distribution, for details see \citep{Zuo1999}.} around a point $\theta$, that ought to be also the depth centre; further, the depth should take higher values near the centre than towards the border.
Properties (P2), (P3) and (P4) in the following definition are inspired precisely by this idea, while (P1) guarantees that common transformations (like standardisation) do not affect the depth of points.

\begin{definition}[Depth function]\label{def:depth}
Let $\textbf{x} \in \mathbb{R}^p$ be a point and $F_{\textbf{X}} \in \mathcal{F}$, the probability distribution function of a random vector $\textbf{X}$ on the Borel $\sigma-$algebra $\mathcal{B}(\mathbb{R}^p)$.
Furthermore, let the mapping $D:~\R^p\times\mathcal{F}\to\mathbb{R}$ be bounded, non-negative and satisfy
\begin{itemize}
\item[P1)] $D(A\textbf{x} + \textbf{b}, F_{A\textbf{X}+\textbf{b}}) = D(\textbf{x}, F_{\textbf{X}})$ holds for any random vector $\textbf{X} \in \R^p$, any $p\times p$ non-singular matrix $A$ and any vector $\textbf{b}\in\R^p$;
\item[P2)] $D(\boldsymbol{\theta}, F_{\textbf{X}}) = \sup\limits_{\textbf{x} \in \R^p} D(\textbf{x}, F_{\textbf{X}})$ holds for any $F_{\textbf{X}} \in \mathcal{F}$ having centre $\boldsymbol{\theta}$;
\item[P3)] for any $F_{\textbf{X}} \in \mathcal{F}$ having deepest point $\boldsymbol{\theta}$, $D(\textbf{x}, F_{\textbf{X}}) \leq D(\boldsymbol{\theta} + \alpha(\textbf{x} - \boldsymbol{\theta}), F_{\textbf{X}})$ holds for any $0 \leq \alpha \leq 1$;
\item[P4)] $D(\textbf{x}, F_{\textbf{X}}) \to 0$ as $\norm{\textbf{x}} \to \infty$ for each $F_{\textbf{X}} \in \mathcal{F}$.
\end{itemize}
Then $D(\cdot, F_{\textbf{X}})$ is called a \textit{statistical depth function}.
\end{definition}

A last key feature of depths, is the identification of a unique centre, regardless of the distribution being uni- or multi-modal. The reason why we emphasise it will be clear in Subsection~\ref{ssec:network-depth}.

Currently, there exist many multivariate depth functions, fulfilling all or just some properties of the definition \ref{def:depth}, differing in robustness to outliers, ease of interpretation, and computational complexity among others.
The halfspace Tukey depth is one of the most famous and well studied multivariate depth functions.
Given that it is based on the ``projection-pursuit'' methodology, it behaves well overall compared with its competitors.
Unfortunately, its exact computation is challenging (or prohibitive) in high-dimensional spaces.
Indeed, given a point $\textbf{x} \in \R^p$ and a p-variate random sample $(\textbf{X}_1, \dots, \textbf{X}_n)$ with empirical distribution $\hat{F}_n$, the (empirical) halfspace depth -- $\HDe_n(\textbf{x}, \hat{F}_n) = \min \left\{\sum\limits_{i=1}^n \textbf{u}^t \textbf{X}_i \leq \textbf{u}^t \textbf{x} : \norm{\textbf{u}} = 1 \right\}$ -- requires to consider projections along all possible $\textbf{u} \in \mathbb{R}^p$ such that $\norm{\textbf{u}}=1$.

Our new depth is based on the original idea of Tukey and it addresses the main drawback of $\HDe$: the high computational cost.

\subsection{Projected Tukey depth} \label{ssec:ptd}

\begin{definition}[Projected Tukey depth]
Let $\textbf{X}$ be a random vector in $\R^p$ with distribution $F_\textbf{X}$, $\textbf{X}_1, \textbf{X}_2$ being two independent copies and $\Sigma$ the covariance matrix\footnote{Assume that $\Sigma$ is invertible or use Moore-Penrose pseudo-inverse.} of $\textbf{X}$ (i.e. assume $\textbf{X} \in L^2(\R^p)$).
For a given $\textbf{x} \in \mathbb R^p$ define the univariate random variable
\begin{equation} \label{eq:Y-x}
Y_x = (\textbf{X}_1 - \textbf{x})^t \Sigma^{-1} (\textbf{X}_2 - \textbf{x})
\end{equation}
then,
\begin{equation}
\PTD(\textbf{x}, F_{\textbf{X}}) = \HDe(0, F_{Y_x}).
\end{equation}
\end{definition}

\begin{theorem}
  The Projected Tukey depth is a statistical depth function.
\end{theorem}

\proof
In the proof an equivalent\footnote{Equivalence can be easily proved writing the probability space $(\Omega, \mathcal{E}, \mathbb{P})$ and $\textbf{X}:\Omega \to \mathbb{R}^p$ with associated probability measure $\mu_X(I) = \mathbb{P}(\textbf{X}^{-1}(I))$ for each Borel set $I \subset \R^p$. Then, $\nu = \mu_{X_1} \otimes \mu_{X_2}$ is the product measure associated to $(\textbf{X}_1, \textbf{X}_2)$. In the following we write $\mathbb{P}(H)$ instead of $\mu_X(H)$ for $H \subset \R^p$, with a negligible ambiguity, and $F_{\textbf{X}}(\textbf{y})=\mathbb{P}(X_1 \leq y_1, \dots, X_p \leq y_p)$ for $\textbf{y}=(y_1, \dots, y_p)\in\mathbb{R}^p$.}
definition is also used: for each realisation $\textbf{x}_2 = \textbf{X}_2(\omega)$ define the univariate random variable
\begin{equation} \label{eq:Y-x-x2}
Y_x| x_2 = (\textbf{X}_1 - \textbf{x})^t \Sigma^{-1} (\textbf{x}_2 - \textbf{x})
\end{equation}
and denote by $Y_x| X_2 = \{ (\textbf{X}_1 - \textbf{x})^t \Sigma^{-1} (\textbf{x}_2 - x) : \textbf{x}_2 \in \textbf{X}_2(\Omega)\}$ the family of random variables conditioned on $\textbf{X}_2$.
Then the Projected Tukey depth of $\textbf{x}$ w.r.t. $F_{\textbf{X}}$ is
\begin{equation}
\PTD(\textbf{x}, F_{\textbf{X}}) = \inf\left\{ \HDe(0, F_{Y_x|x_2}) : \textbf{x}_2 \in \textbf{X}_2(\Omega) \right\} \stackrel{not.}{=} \inf\left\{ \HDe(0, F_{Y_x|X_2}) \right\}.
\end{equation}
Clearly $\PTD$ is bounded and non-negative.
\begin{itemize}
\item[P1)] Affine invariance: we have to prove that for any non singular matrix $A \in \R^{p\times p}$ and any vector $\textbf{b} \in \R^p$
\[
\PTD(A\textbf{x} + \textbf{b}, F_{A\textbf{X} + \textbf{b}}) = \PTD(\textbf{x}, F_{\textbf{X}}). \\
\]
The covariance matrix of $A\textbf{X} + \textbf{b}$ is $A^t \Sigma A$ so we have
\[
\begin{aligned}
Y_{A \textbf{X} + \textbf{b}} & = \left(A \textbf{X}_1+ \textbf{b} - (A \textbf{X} + \textbf{b})\right)^t \left( A^{-1} \right)^t \Sigma^{-1} A^{-1} \left(A \textbf{X}_2 + \textbf{b} - (A \textbf{X} + \textbf{b})\right) \\
& = \left(A(\textbf{X}_1 - \textbf{x})\right)^t \left( A^{-1} \right)^t \Sigma^{-1} A^{-1} \left(A(\textbf{X}_2 - \textbf{x})\right) = Y_x
\end{aligned}
\]
hence $Y_{A \textbf{X} + \textbf{b}} \stackrel{d}{=} Y_\textbf{X}$; the thesis follows since property P1 holds for $\HDe$.\\
From now on we assume w.l.o.g. that $\Sigma = \text{Id}$.
\item[P2)] The proof of maximality at centre property involves probability distributions that are \textit{symmetric} with around $\boldsymbol{\theta}$.
There are several definitions of multivariate symmetry: spherical (most restrictive), elliptical, angular and halfspace (less restrictive) \citep{Zuo1999}; P2 can be proven in the latter, general case.
Let the distribution of $\textbf{X}$ be $H$-symmetric about a unique centre $\boldsymbol{\theta}$, i.e. in terms of the underlying probability measure $\mathbb{P}(H_{\theta}) \geq \frac12$ for every closed halfspace $H_{\theta}$ containing $\boldsymbol{\theta}$ on its border.
Assume $\boldsymbol{\theta} = 0$ thanks to P1.
We ask if $\PTD(0, F_{\textbf{X}}) \geq \PTD(\textbf{x}, F_{\textbf{X}})$ for all $\textbf{x} \in \mathbb R^p$. From \eqref{eq:Y-x-x2} and \citep[Th.2.3.2.]{Zuo1999}
\begin{equation*}
  \PTD(0, F_{\textbf{X}}) = \inf \left\{ \HDe\left(0, F_{Y_{0}|X_2}\right) \right\} \geq \frac12
\end{equation*}
similarly each projection $\Pi_{\textbf{x}, \textbf{x}_2}(0) = (0 - \textbf{x})^t(\textbf{x}_2 - \textbf{x})$ on the line through $\textbf{x}$ and $\textbf{x}_2$ is such that $\HDe\left(\Pi_{\textbf{x}, \textbf{x}_2}(0), F_{Y_{x}|x_2}\right) \geq \frac12$, hence $\inf \left\{ \HDe\left(0, F_{Y_{0}|X_2}\right) \right\} \geq \frac12$.
Since the halfspace depth $\HDe$ satisfies the maximality at centre property $\HDe\left(\Pi_{x, x_2}(0), F_{Y_{x}|x_2}\right) \geq \HDe\left(0, F_{Y_{x}|X_2}\right)$ and $\HDe\left(0, F_{Y_{x}|X_2}\right) \leq \frac12$ for the uniqueness of the centre of symmetry assumption, it cannot happen that $\PTD(x, F_{\textbf{X}}) > \frac12$.
\item[P3)] Monotonicity relative to deepest point: assuming w.l.o.g. $\boldsymbol{\theta} = 0$ we have to prove that $\PTD(\textbf{x}, F_{\textbf{X}}) \leq \PTD(\alpha \textbf{x}, F_{\textbf{X}})$ for $\alpha \in [0, 1]$.
The interesting case is when $\alpha \in (0, 1)$.
The problem is that we have to compare two families of marginals $Y_x| X_2 = \{ (\textbf{X}_1 - \textbf{x})^t \Sigma^{-1} (\textbf{x}_2 - \textbf{x}) : \textbf{x}_2 \in \textbf{X}_2(\Omega)\}$ and $Y_{\alpha x}| X_2 = \{ (\textbf{X}_1 - \alpha \textbf{x})^t \Sigma^{-1} (\textbf{x}_2 - \alpha \textbf{x}) : \textbf{x}_2 \in X_2(\Omega)\}$.
P3) has been proved only for spherical and elliptical symmetric distributions.
In the first case we use a characterisation of spherical symmetry stating that rotations around the centre do not affect the distribution, formally $\textbf{X} \stackrel{d}{=} A \textbf{X}$ for orthogonal matrices $A$. Therefore, marginals on lines through the centre are all equal and the thesis follows immediately.
Elliptic symmetry reduces to spherical when $\Sigma = \text{Id}$ so by means of P1) we ho back to previous case.
\item[P4)] Vanishing at infinity -- $\PTD(\textbf{x}, F_{\textbf{X}}) \to 0 $ as $\norm{\textbf{x}} \to \infty?$\\
Assume w.l.o.g. $\mathbb{E}(\textbf{X}) = 0, \Sigma = \text{Id}$.
Take $\textbf{x}_1, \textbf{x}_2$ fixed, then $y_x = \norm{\textbf{x}_1 - \textbf{x}}\norm{\textbf{x}_2 - \textbf{x}} \cos{\phi}$ where $\phi$ is the angle between the two vectors $\textbf{x}_i - \textbf{x}$, $i = 1, 2$. If $\phi \in \left(\frac{\pi}{2}, \frac{3\pi}{2}\right)$ then $y_x \leq 0$, otherwise $y_x \geq 0$.

We want to show that for $\textbf{x}$ sufficiently far away from the deepest point $\cos(\phi) > 0$ so that $y_x \to \infty$ for each realisation. Then $\HDe(0, F_{Y_x}) \to 0$.

If $\textbf{X}$ is bounded in $\R^p$, as soon as $\textbf{x} \in \R^p$ is outside the ball of finite radius containing $\textbf{X}(\Omega)$ the angle $\phi < \frac{\pi}{2}$ and $y_x \geq 0$.

If $\textbf{X}$ is not bounded in $\R^p$, using Chebyshev's inequality on the univariate r.v. $Y_{x|x_2}$, i.e. $\forall \epsilon > 0$
\begin{equation*}
  P\left( \left| Y_{x|x_2} \right| \geq \epsilon \right) \leq \frac{1}{\epsilon^2}
\end{equation*}
we have that the probability of finding a point arbitrarily distant from $\mathbb{E}(\textbf{X})$ is small. For $\textbf{x}$ such that $\norm{\textbf{x}} \to \infty$ we can then conclude that $P(\phi < \frac{\pi}{2}) \to 1$ and the thesis follow.
\end{itemize}
This closes the proof. \qed

The sample version of the $\PTD$ is built around \eqref{eq:Y-x-x2}. It is similar to $\HDe_n$ with two important differences: first, only a subset of possible projecting directions are considered and exactly those indicated by that; secondly we do not take orthogonal but variance-scaled projections, akin in spirit to Mahalanobis distance.

\begin{definition}[Sample PTD]
Let $(\textbf{x}_1, \dots, \textbf{x}_n)$ be a random sample and $\hat{F}_n$ its empirical distribution; $\forall i, j = 1, \cdots, n$ and $\textbf{x} \in \R^p$
\begin{align}
y_{ij} & = (\textbf{x}_i - \textbf{x})^t \hat{\Sigma}^{-1} (\textbf{x}_j - \textbf{x}) \label{eq:ptd-rescale}
\end{align}
is the projection of the $i$--th sample point onto the line through $\textbf{x}$ and $\textbf{x}_j$. Collecting them for all $j$ we get a univariate sample
\begin{align*}
  y_{\cdot j} & = (y_{1j}, \cdots, y_{nj})
\end{align*}
and we take the minimum probability mass carried by \textit{halflines} with $0$ on the boundary
\begin{align}
PTD_n(\textbf{x}, \hat{F}_n) & = \min \left\{ \HDe_n(0, \hat{F}_{y_{\cdot j}}) : j = 1, \dots, n \right\} \nonumber\\
& =  \min \left \{\min \left( \left|\{i : y_{ij} \leq 0, i \in 1\dots n\}\right|, \left|\{i : y_{ij} \geq 0, i \in 1\dots n\} \right| \right) : j = 1, \dots, n \right\} \label{eq:sample-ptd}.
\end{align}
In the following we denote $PTD_n$ again by $PTD$.\\
In matrix form \eqref{eq:ptd-rescale} becomes $\tilde{X} \hat{\Sigma}^{-1} \tilde{X}^t$, where $\tilde{X}$ is obtained subtracting $\textbf{x}$ from each row of $X$.
\end{definition}

The set $\{\PTD(\textbf{x}, F_{\textbf{X}}),~ \forall \textbf{x} \in \textbf{X}(\Omega)\}$ is called the \text{depth space} of $\textbf{X}$, or associated with distribution $F_{\textbf{X}}$.
In the empirical case the depth space is a vector whose entries are the depth value for each observation and this will be the centrality vector $C_{PTD}(G)$ of nodes in graph $G$.
Another useful family of sets are the so-called depth trimmed (or central) regions $R_{\alpha}$, which generalise quantile-intervals, given $\alpha \in [0, 1]$ the depth region of order $\alpha$ is the set characterised by
\begin{equation}\label{eq:ptd-region}
  \mathbb{P}\left( \{ \textbf{x} \in \R^p : \PTD(\textbf{x}, \mathbb{P}) \geq q_{\alpha} \} \right) = 1 - \alpha.
\end{equation}
For instance, $R_{\alpha = 0.99}$ is the depth central region containing the 1\% deepest points.
The set $C_{\alpha}\{ \textbf{x} \in \R^p : \PTD(\textbf{x}, \mathbb{P}) = q_{\alpha} \}$ is called depth contour of depth $q_{\alpha}$.
The empirical version of the contour set is computed as the convex hull containing the most central fraction $1 - \alpha$ of sample points and it is denoted by $C_{\alpha, n}$; $q_{\alpha}$ becomes $q_{\alpha, n}$ and the depth region $R_{\alpha, n}$. Again, as we will always deal with empirical quantities, we drop the subscript $n$ in the notation.

The Projected Tukey depth can be computed exactly also in high-dimensional spaces. Given $n$ points in $\mathbb{R}^p$, i.e. $X \in \mathbb{R}^{n \times p}$, the algorithm of the PTD requires (i) the inversion of its sample covariance matrix $\hat{\Sigma} = cov(X)$, which costs $O(p^3)$ in the worst case and (ii) the evaluation of the matrix product $\tilde{X} \hat{\Sigma}^{-1} \tilde{X}^t$, which has complexity $O(n p^2) + O(p n^2)$ using the naive matrix multiplication algorithm. Usually $p$ is not very high since we use dimensionality reduction techniques to embed our networks in space, so in the end the evaluation of the PTD of a point w.r.t. a cloud of $n$ points has a complexity of $O(pn^2)$ and of $O(pn^3)$ for the whole data cloud.

In the next Subsection we show how this multivariate depth function can be used to generalise the definition of median to complex networks.

\subsection{The Network Depth: a New Centrality}\label{ssec:network-depth}

\begin{figure}
  \centering
  \includegraphics[width=\textwidth]{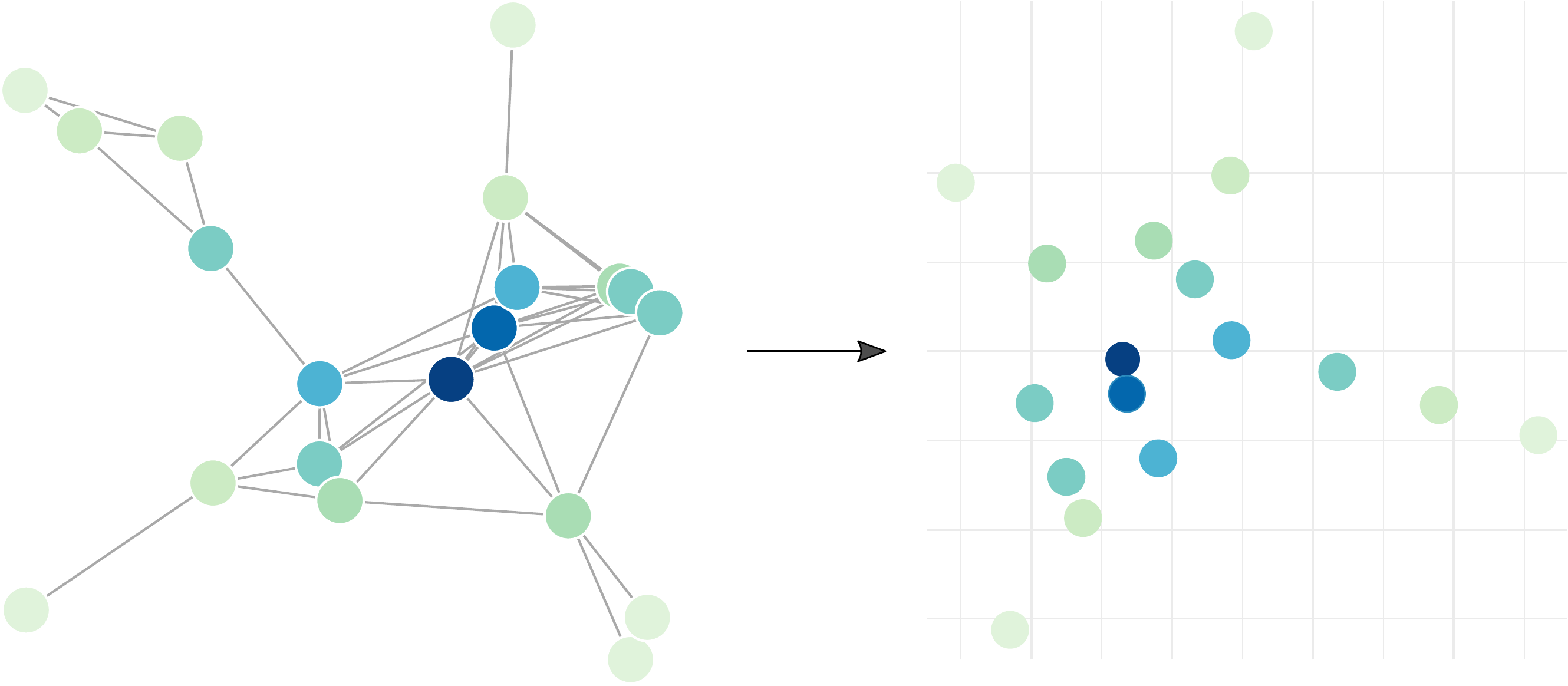}
  \caption{Moving the node centrality problem to that of finding the depth space of the multivariate distribution associated to the network.}
\end{figure}

The key step linking the Projected Tukey depth to network centralities is network embedding.
A network $G = (V, E)$ embedded in some space $\R^p$ is a cloud of points $X(G)$ distributed accordingly to the connectivity of $G$.
We map the problem of finding the centrality $C(i, G)$ of the node $i$, to the problem of computing the depth $PTD(\textbf{x}_i, \hat{F}_G)$ of the point $\textbf{x}_i$ w.r.t. the (empirical) distribution $\hat{F}_G$ of $X(G)$.
The idea is that $\hat{F}_G$ reflects the connectivity of graph $G$, adjacent vertices should lie nearer than nodes that do not share edges.
Thus, the priority for our embedding is preserving distances.

On an undirected weighted connected network, we have a distance given by length of shortest-paths, $D_{sp}$.
If the graph is directed, $D_{sp}$ is no longer a metric and a non-metric embedding may be needed.
Another distance can always be defined on (connected) networks, that is based on diffusive random walks, the diffusion distance $D_t$ \citep{DeDomenico2017}.
Let $\textbf{p}(t | i)$ be the vector with entries $p_k(t | i)$ representing the probability that a random walker starting at $t=0$ in node $i$ is found in $k$ at time $t$. It is well known that this probability is linked to the network structure, in particular given the (random walk) normalised Laplacian $\tilde{\textbf{L}} = \text{Id} - D^{-1} A$, where $D$ is the diagonal matrix of node degrees, the master equation of the ``continuised'' random walk is
\begin{equation*}
  \dot{\textbf{p}}(t) = - \textbf{p}(t) \tilde{\textbf{L}}.
\end{equation*}
The existence of a mapping from $G$ to the geometric space induced by Markov dynamics, makes it possible to define the diffusion distance at time $t$, as shown in \citep{Coifman2005} and \citep{DeDomenico2017}
\begin{equation}
  D_t(i, j; G) = \norm{\textbf{p}(t | i) - \textbf{p}(t | j)}_2
\end{equation}
which expresses how nodes $i$ and $j$ are efficient in exchanging information.
In other terms, $D_t(i, j; G)$ is small if there is a large transition probability from $i$ to $j$ along walks of length at most $t$.
Observe that $D_t$ does not assume that the spreading takes place through shortest-paths and it is robust w.r.t. small fluctuations in the structure of the network.
If the network is not connected, one has to regard each component separately or can consider a more general random walk dynamics with a teleportation term, as in diffusive processes used by PageRank \citep{Page1999} and Infomap \citep{Rosvall2008}.

Given a distance matrix describing structural and/or functional properties of our network we can recover (possibly with some approximations) the position of the $N = |V|$ nodes in a space of dimension $p \in \mathbb{N}$. $p$ is always a free parameter of network embedding procedures.
We call \textit{perfect embedding} a configuration $X \in \mathbb{R}^{N \times p}$ of $N$ points in $\mathbb{R}^p$ for an appropriate $p \geq 1$ such that $\norm{\textbf{x}_i - \textbf{x}_j}_2 = D(i, j)$ for all $i, j = 1, \dots, N$, where $\textbf{x}_i$ is the $i$--th row of matrix configuration $X(G)$ and corresponds to the coordinates of node $i$ in $\R^p$.
For the moment, let us assume that, for a given distance matrix $D$, we have some $p$, such that the embedding of network $G = (V, E)$ in $\R^p$ is perfect (perfect embedding assumption).
In statistical terms we do not lose information passing from $G$ to $X = X(G)$ and the centrality problem is equivalent to finding the depth space of $X(G)$.

Let us denote by $\PTD(v; G, D_{sp}, p)$ the PTD centrality of node $v \in G$ with respect to geodesic embedding of $G$ in $\R^p$ and by $\PTD(G; D_{sp}, p)$ the corresponding depth space, i.e. the network depth centrality vector with parameters $D_{sp}, p)$. We do not use the initial notation $C_{PTD}$ because the network depth, through the embedding step, depends on some free parameters.
Similarly $\PTD(v; G, D_{t}, t, p)$ is the depth of node $v$ in the diffusion embedding with parameters $(t, p)$ and $\PTD(G; D_{t}, t, p)$ refers to the centrality vector $\forall v \in G$.
If not ambiguous $G$ is omitted.

Using Borgatti's notation, the matrix $W$, containing the information on connectedness/social proximity, in our case is $D$ or equivalently (under our perfect embedding assumption) $X$.
Then the PTD is exactly the statistic summarising each node's share of total cohesion, as a matter of fact $T(W) = f \left( \tilde{X}\Sigma^{-1} \tilde{X} \right)$.
Taking the Projected Tukey depth as a Summary Type makes the centrality, not only robust to outliers, but it also enables the extension to structured data of nonparametric statistical tools based on order statistics (e.g. location and scale estimators).

The other three features characterising centralities are Walk Type, Walk Property and Walk Position.
Independently on the metric, $\PTD$ is a radial measure and, since it ignores multi-modality of distributions, it is not intended for the identification of local centres in a multi-community structure.

\begin{figure}[!h]
\centering
\subfloat[1D]{\includegraphics[width=.45\textwidth]{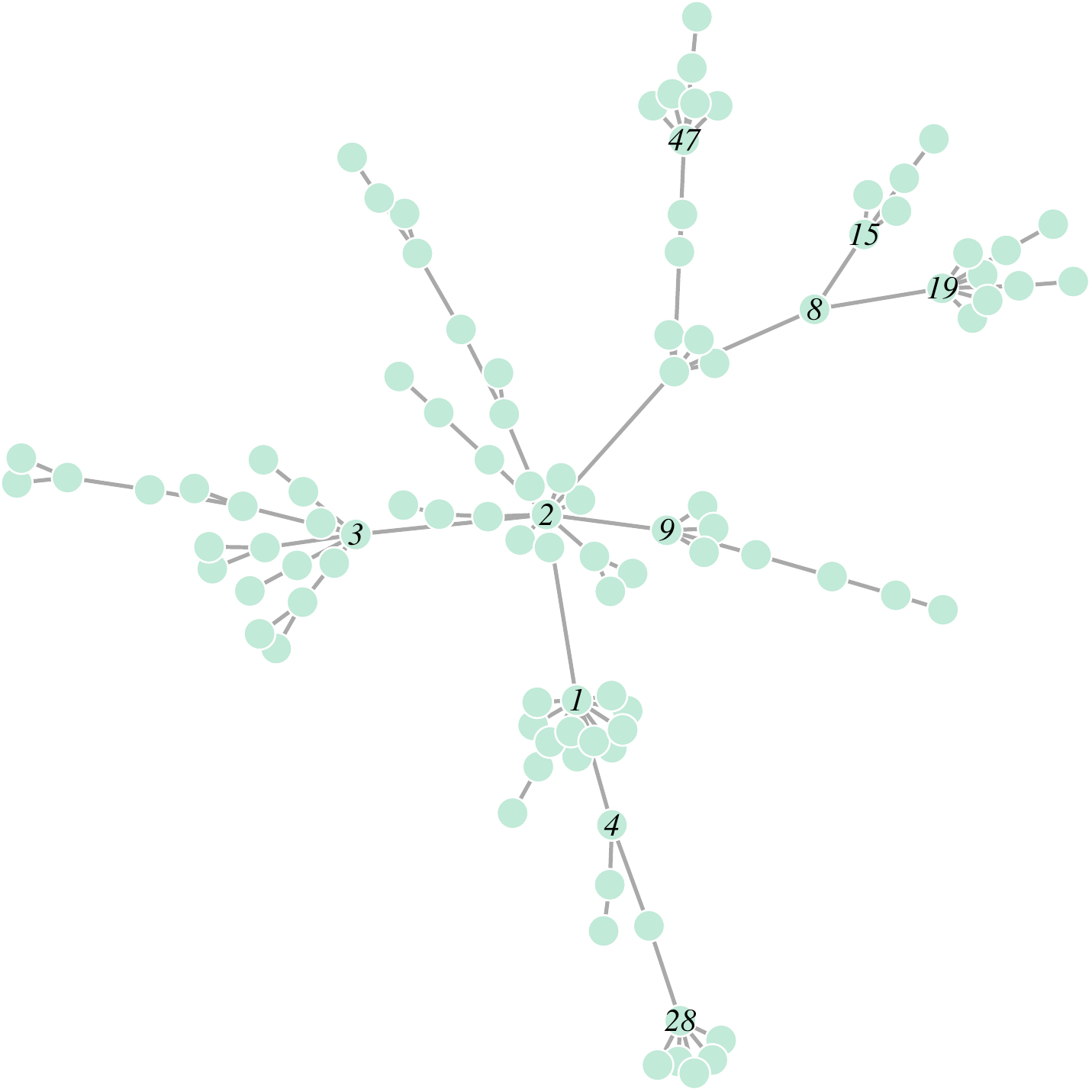}}
\subfloat[2D]{\includegraphics[width=.45\textwidth]{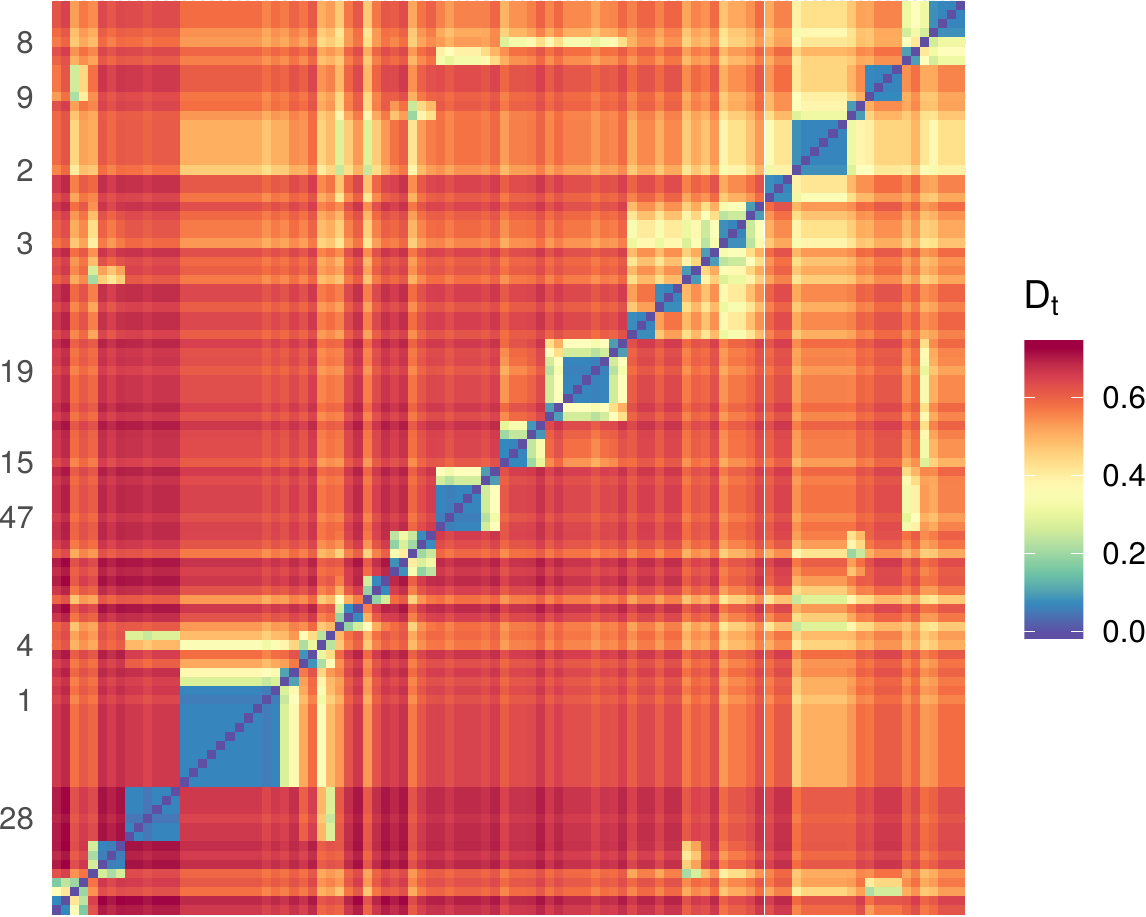}}
\caption{A Barabasi-Albert network with 100 nodes and its $D_t$ diffusion distance matrix for $t=3$. The blocks in the distance matrix clearly reflect the groups in the network. Consider for instance $v_1$, although it is linked to $v_2$, which is a hub of the network, a random walker starting in $v_1$ will spend some time trapped in its branch of the network.}
\label{fig:dist-mat}
\end{figure}

On the contrary Walk Type and Property depend on the distance.
$\PTD(v; G, D_{sp}, p)$, like Freeman's closeness and betweenness, takes into account only geodesics.
In terms of walks, these measures assume the walker has a global knowledge of the network and it navigates it through the shortest possible paths\footnote{We do not only have a constraint on length, but also on the walks, indeed paths are sequences of unrepeated edges and nodes.}.
Finally, $\PTD(v; G, D_{sp}, p)$ is a length measure (Walk Property) since $\PTD(x, F_{\textbf{X}})$ in $\R^p$ can be interpreted as the ``tailness'' of point $x$ with respect to the distribution $F_{\textbf{X}}$.
In other words, a point which lies deep inside the data cloud in the $D_{sp}$ embedding has small values of geodesic distance with all other points.
On the other side, $\PTD(v; G, D_{t}, t, p)$ summarises structural information gathered through random walk exploration of the network.
The diffusion time $t$ sets a constraint on the length of otherwise unrestricted edges' sequences.
As observed in \citep{DeDomenico2017}, $t$ plays the role of a scale parameter: for small values of $t$ the micro-scale structure is revealed, see Fig.~\ref{fig:dist-mat}.
In this case the network is embedded in a high dimensional space and the data cloud is very sparse with nodes with smaller $D_t$ lying deeper than all others, which are placed on a convex hull and have a depth of $\frac1N$.
When $t$ increases diffusion distances display two different behaviours: $D_t$ among nodes exchanging efficiently information tends rapidly to zero, whereas $D_t(u, v)$ shrinks slowly if there is a small probability that two random walkers starting in $u$ and $v$ respectively meet somewhere in the network, following paths of length at most $t$.
Since the Projected Tukey depth is not affected by the magnitude of the distances, but only on the relative position of points in space, hence after a certain amount of time, $t > t^*(G)$ the network depth spans a larger range and $\PTD(v; G, D_{t}, t, p)$ remains stable.
Of course, in the limit for $t \to \infty$, $D_t \to 0$ for all pairs of nodes, here small fluctuations due to non-perfect embedding might greatly affect the depth of points.

Finally, the Walk Property of the depth w.r.t. diffusion embedding is clearly volume, as a matter of fact $i$ and $j$ are near in the diffusion manifold if there is a high probability that two random walkers starting respectively in $i$ and $j$ meet somewhere in the network. In terms of connections, $D_t(i, j)$ is small if there are a lot of short paths connecting them.

The perfect embedding assumption is satisfied only if the original node distances $D(i, j)=D_E(i, j)$ are Euclidean distances in $\R^p$, then $D_E(i, j) = \norm{\textbf{x}_i - \textbf{x}_j}$ and the configuration matrix $X$ is reconstructed without errors.
In general, $\norm{\textbf{x}_i - \textbf{x}_j} \approx D(i, j)$ and matrix $X$ is the result of a possibly non-metric Multidimensional Scaling (MDS).
It is also not mandatory to have a distance matrix since MDS deals with general dissimilarity matrices $\Delta = \{\delta_{ij}\}_{i, j}$.
In the next subsection we will see the embedding step more in detail.

\subsection{Network embedding}\label{ssec:net-emb}

Multidimensional scaling is a collection of techniques aiming to find structures in multidimensional data, given a proximity matrix $\Delta$ of observations and an embedding dimension $p$ \citep{Cox2000}.
The MDS problem can be solved analytically or using iterative procedures, like the majorisation algorithm.
The latter yields usually better results than the analytical solution, although its iterative nature may be time consuming for large $N$.
When solving MDS by means of the majorization algorithm the loss function to be minimised is Kruskal's \textit{stress-1}: given a configuration $X$ of $N$ points in $\mathbb{R}^p$ with $1 \leq p \leq N-1$
\begin{equation}\label{eq:stress-1}
  \sigma(X) = \sum_i \sum_j \left( \delta_{ij} - d_{ij}(X) \right)^2
\end{equation}
where $\delta_{ij}$ are the original distances.
If $\Delta$ is not a proper distance matrix, $\delta_{ij}$ are called dissimilarities and they may need to be transformed into $\hat{d}_{ij}=f(\delta_{ij})$ before the scaling, for instance through a strictly increasing monotonic function $f$ such that order is preserved $\delta_{ij} < \delta_{kl} \Rightarrow f(\delta_{ij}) < f(\delta_{kl})$.
This case falls under the non-metric MDS methods.
Furthermore, unfolding models allow us to deal with non-symmetric matrices $\Delta$ (which may be useful if in- and out-going geodesics have to be considered).

The dimension $p$ of the Euclidean space to which we map the network is then a free parameter of the network depth.
If $p$ is very small compared to $N$, the poor quality embedding will affect the depth too.
Instead, if $p$ is very large, the statistical information of the sample is so diluted in the high dimensional space, that all nodes happen to have the same depth of $\frac1N$, i.e. they lie on a convex hull.
Take for instance a star, a simple planar graph and the geodesic distance $D_{sp}$.
We may think that $p=2$ should be the right embedding dimension but if we perform a metric MDS without any constraint we end up with a large stress value -- since every permutation of nodes different from the star centre are equivalent -- and a configuration $X$ failing to match our expectations.
A better approach is to use a constrained MDS on a 2D-sphere, maybe relaxing also the convergence tolerance.

In general, the most appropriate $p$ is not known a-priori and we end up with an embedding for each $p \in \{1\leq p_{min}, \dots, p_{max}\leq N-1\}$ so that $p$ is a free parameter for the depth.
First of all, embeddings with ``high'' stress-1 values should be removed. What is ``high'' depends on the data.
A typically approach is to look for the ``legendary statistical elbow'' in the stress curve (scree-plot) or to follow Kruskal's Rule of Thumb \citep{Kruskal1964} stating that stress should be less than $0.2$.
Whether possible, it is warmly suggested -- see \citep{deLeeuw2009} and \citep{Mair2016} -- to perform some additional goodness-of-fit analysis, like a permutation test to get a p-value associated with the observed stress and diagnostic plots, e.g \textit{Stress-per-point} plot to see if a subset of points accounts for large values of $\sigma(X)$.
As we already pointed out, for $p \to N-1$ the depth spans an increasingly small range since the sample of $N$ points becomes more and more sparse in $\mathbb{R}^p$, hence the principle of parsimony plays a role in determining $p$ too.
Finally, as we will see in the next section there is an interval between $p_{min}$ and $p_{max}$ where the depth pattern (consequently the ranking) remains stable so that the PTD centrality score vector can be chosen among a $\{\PTD(D, p)\}$ in that interval. With the expression \textit{depth pattern} we mean the ordering of nodes as a function of one free parameter.
A depth pattern plot has the free parameter, e.g. $t$ on the x-axis and the values of $\PTD(v, G; D, t)$ for each node $v\in G$ on the y-axis. This plot enables us to graphically see how the depth range evolves, if the top ranking nodes remain approximately the same (\textit{stable pattern}) or if the ranking changes considerably, indicating noise in the embedding.

The network depth $\PTD(D_t, \cdot, \cdot)$ w.r.t. diffusion distance depends also on $t$ and the scree-plot becomes a heatmap representing the stress surface, Fig.\ref{fig:emb-stress-karate}, still, the same considerations apply for $p$, while $t$ makes the $\PTD(D_t, \cdot, \cdot)$ a multi-resolution centrality measure, as discussed in \ref{ssec:network-depth}.
To obtain a parameter-free centrality without having to fix the values for $p$ and/or $t$, simply integrate over the parameter space -- possibly removing those parameters leading to the worst embeddings -- and look at the aggregate network depth.
In the case of diffusion embedding one may also consider the average diffusion distance matrix, as in \citep{DeDomenico2017}, getting rid of one parameter.

We summarise the main steps the procedure as follows:

\begin{itemize}
  \item[0)] Choose a network metric, remembering that the shortest-path distance provides results which are similar to the closeness centrality and that if you choose the diffusion distance the diffusion time $t$ plays the role of a scale parameter. Compute the distance matrix $D$ (or set of matrices $\{D_t\}_{t\geq 0}$).
  \item[1)] Embed the network $G=(V,E)$ in $\R^p$, using MDS (either solving the scaling problem analytically or through the majorisation algorithm); these statistical methods are preferred, since they give a feedback on the amount of information preserved/lost in the scaling process. Field-knowledge may suggest a dimension $p$, a-priori. As a rule of thumb, for graphs with more than 100 nodes, avoid $p < 4$, remember that $p$ depends on the MDS algorithm one is using.
  \item[2)] For each space configuration of points $X(G) \in \mathbb{R}^{N \times p}$, compute Kruskal's stress function \eqref{eq:stress-1}. It will be useful for selecting a range of reasonable dimensions $p$. It is not mandatory to evaluate the goodness-of-fit of the embedding, but it guarantess you that the subsequent findings are not the result of chance.
  \item[3)] For each space configuration of points $X(G)$, compute the depth space $\PTD(G; D, p)$, i.e. the network depth centrality vector, by means of equations \eqref{eq:ptd-rescale}-\eqref{eq:sample-ptd}. If the depth space reduces to just one value, $\frac1N$, the dimension $p$ is too large, so you can focus on smaller embedding spaces. To rule out values of $p$ which are too small, plot the stress-1 values against $p$ (scree plot), values larger than 0.2 rarely lead to good results. Additionally you can look at high order depth central regions $R_{\alpha}$, if there is a high variability in the nodes belonging to these regions, the depth pattern is surely not stable, so you may think of removing too small dimensions as well.
  \item[4)] ``Stable'' depth patterns represent an indicator for good embedding dimensions $p$, where stable means that there are not nodes making big jumps across the depth ranking. In other terms, high order depth regions contain more or less the same nodes (possibly swapping their order inside the region).  You can either select a single $p$ and have a single depth space or aggregate the depth score vectors, integrating over the (whole or part of) parameter space.
  \item[5)] Given the vector of depth values, quantiles $q_{\alpha}$ are easily computed and so you can find depth central regions $R_{\alpha}$, while the network median is the node with maximum depth.
\end{itemize}

\section{Applications to Empirical Networks}\label{sec:applications}

In this section we apply the network depth to some examples of social, biological and infrastructure networks and compare it with the measures in Tab.~\ref{tab:cms-topology}: degree, eigenvector, closeness, betweenness and distance-weighted fragmentation centrality\footnote{keyplayer R package available on CRAN.}. As already mentioned we compute only these five centralities, since they are representative of other measures and the goal of our work is not to provide another comparative study on centralities applied to real networks, but to propose a statistically grounded centrality measure. The novelty of our work lies in the approach to the well-studied problem of node importance inside a network: the network depth is a generalisation of statistical depth function to structured data, through network embedding. On the other hand, depths induce multivariate order relations and enable the generalisation of order statistics (median and quantiles) to high dimensional spaces.

\subsection{Social Networks}

\subsubsection{Zachary's Karate Club.}

\begin{figure}[t]
\centering
\includegraphics[width=\textwidth]{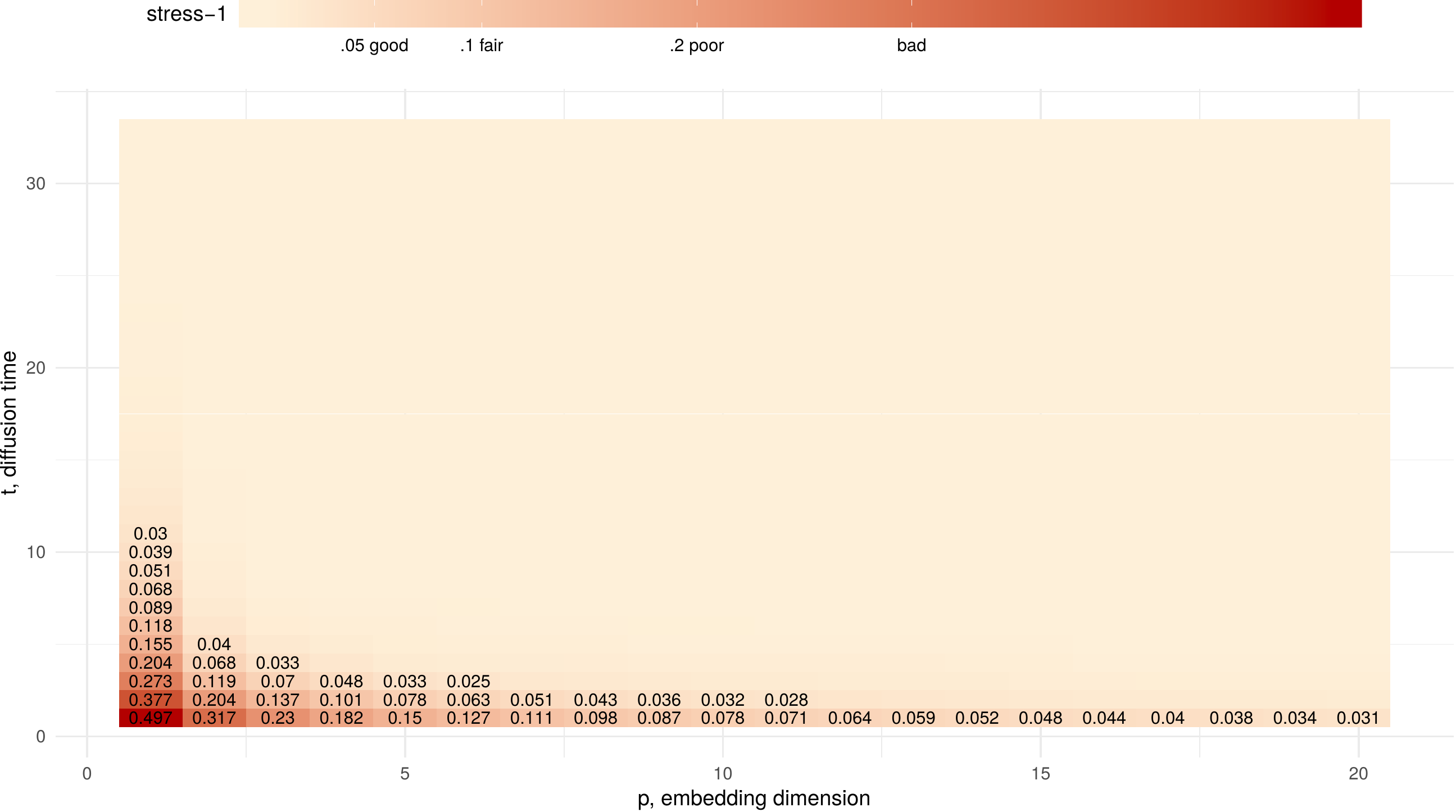}
\caption{Stress surface of the Karate Club Network's diffusion embeddings. Linear MDS places the 34 Actors building the network into $\mathbb{R}^p$ minimising the sum of squared differences (stress-1) between points in the Euclidean space and original diffusion distances $D_t$ on the graph, for varying $p$ ($x-$axis) and $t$ ($y-$axis). Main observations: (i) stress-1 is a decreasing function of $p$, implying that we cannot take it alone as a measure of embedding/centrality quality. (ii) 1D embeddings are poor quality in general, also for small networks.}
\label{fig:emb-stress-karate}
\end{figure}

The first example we propose is the famous Zachary's Karate Club network \citep{Zachary1977}, it is a small ($N=34$) social network where Actors are members of a university karate club and weighted edges represent interactions among them outside the club in a period of three years.
It has a modular structure with two main communities, the \textit{factions} with heads Mr. Hi, the president, and the coach John A, into which the group split after a conflict.
The plot in Fig.~\ref{fig:emb-stress-karate} shows the stress-1, $\sigma$, for embeddings with parameters $(D_t; t, p)$, which is in general low (stress-1 depends also on the sample size $N$, \citep{Mair2016}), with higher values for small $t, p$.
The pattern visible in the heatmap is found also for the other networks we analysed. Diffusion embeddings for short diffusion times tend to have a larger stress value since there are many nodes equally far away from each other according to $D_t$ and these are interchangeable in terms of configuration $X(G)$ (similarly to a star graph). As $p$ increases $\sigma$ decreases, but remember that the sample becomes more and more diluted in high-dimensional spaces.

We then move to analyse the \textit{depth patterns} plotting the depth value of each node as a function of $t$ or $p$.
These plots represent a well suited graphical tool to study the centrality range, scores and ranking, their stability and relation with the free parameters.
Statistics is known to make a massive use of graphics, in particular our idea of depth pattern plots draws inspiration from forward plots in Forward Search analysis for outliers detection, see \citep{Atkinson2010}.
The key idea there is to monitor quantities of interest, e.g. test statistics, as the model is fitted to data subsets of increasing size.
In our case, we study depth values $\PTD(i, G; \dots)$ as space dimension $p$ or diffusion time $t$ increase.

\begin{figure}[t]
\centering
\includegraphics[width=\textwidth]{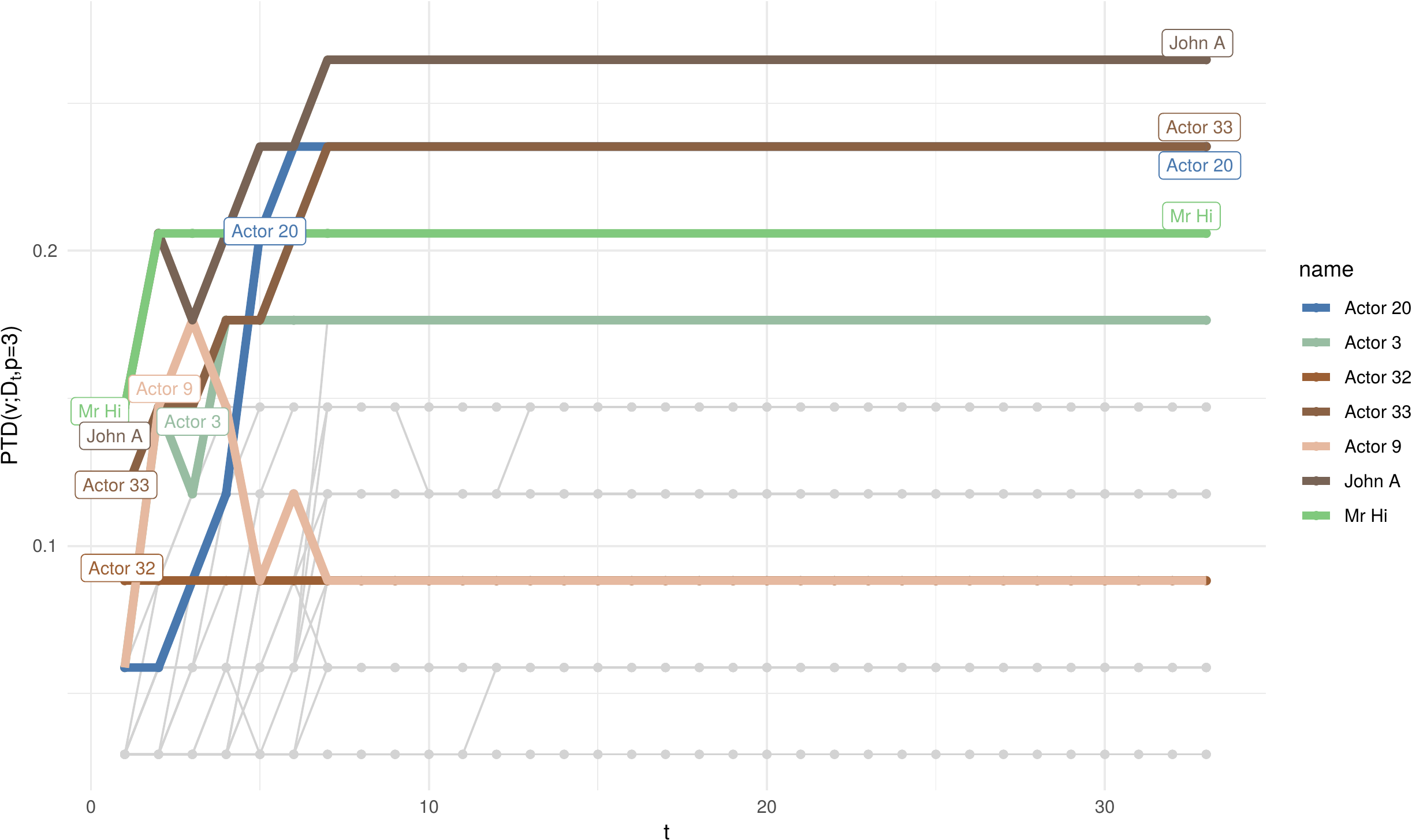}
\caption{Depth patterns plot: depth of Actors in Zachary's network according to $\PTD(D_t, t, p = 3)$, as diffusion time increases. Lines are coloured if, for some $t$, the corresponding Actors belong to the depth trimmed region $R_{\alpha, n}$ \eqref{eq:ptd-region} of order $\alpha = 0.9$, that is if the corresponding Actors have a degree of (depth) centrality of at least 90\% with respect to the distribution induced by the network.}
\label{fig:karate-Dt-ptd-p3}
\end{figure}

\begin{figure}[ht]
\centering
\includegraphics[width=\textwidth]{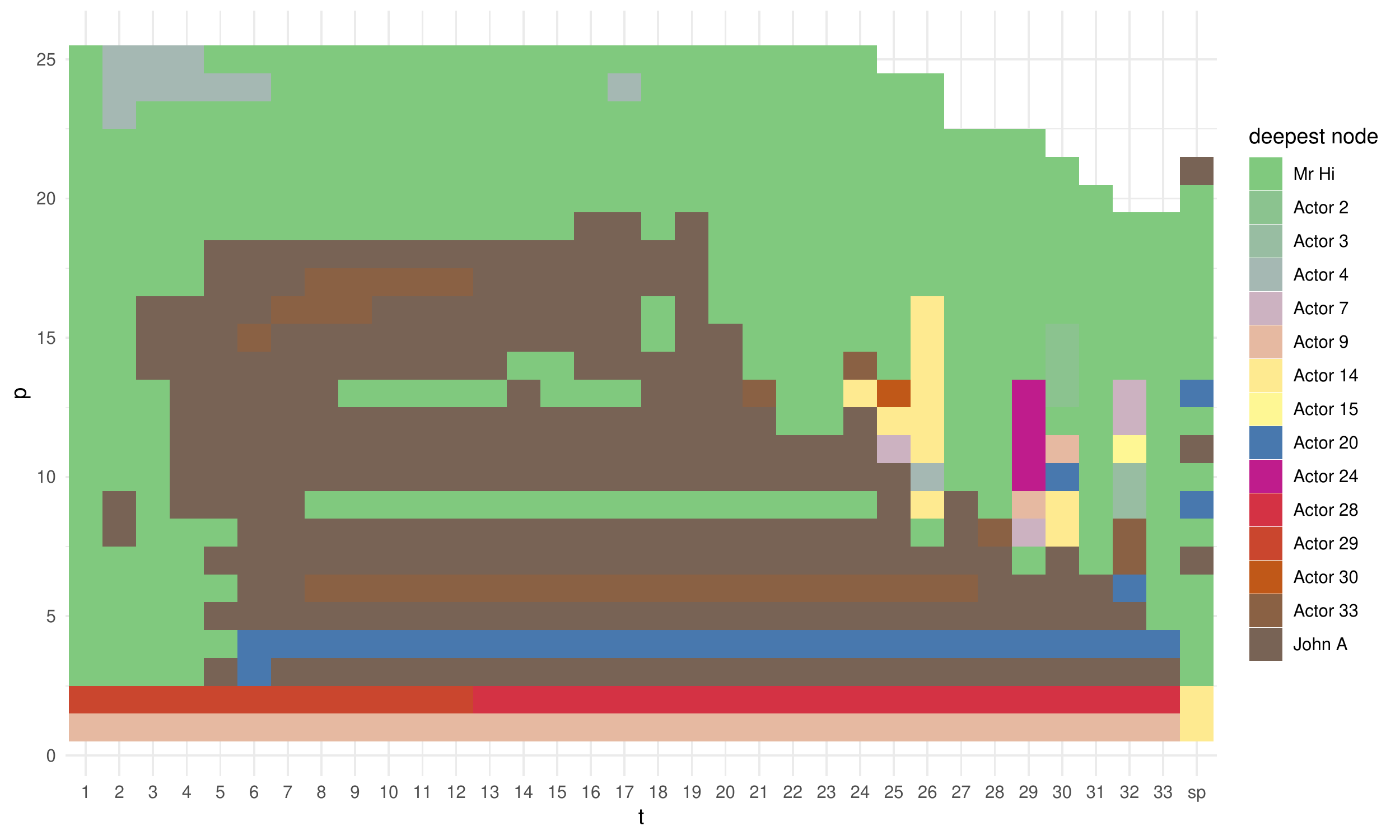}
\caption{Comparing the (generalised) median in Zachary's Karate Club Network across the parameter space. The deepest node (median) w.r.t. $\PTD(D_t, t, p)$ and $\PTD(D_{sp}, p)$ in the right-most column. Mr Hi and John A. are the most central point for a large part of the space parameter, followed by Actors 20, 9 and 33. Actor 20 in particular is also the third most central node according to betweenness and closeness, while Actor 33 scores three in the degree ranking. For small values of $t$, ignoring poor quality embeddings in $\mathbb{R}$ and $\mathbb{R}^2$, the most central vertex is Mr. Hi and the same is also true in the shortest-path embedding spaces. Missing tiles indicate that, in those embeddings, all points lie on a convex hull and it is hence meaningless to compute their depth.}
\label{fig:deepest-karate}
\end{figure}

We can find $\PTD(D_t; t, p = 3)$ as a function of $t$, in Fig.~\ref{fig:karate-Dt-ptd-p3}
For $t=1$ the depth space has a small range $\left[\frac1N, \frac5N\right]$, while for $t\geq 7$ it remains stable, with little variations on depth levels.
Mr. Hi and John A. are the most central nodes until $t=4$, after which Actors 20 and 33 gain the second position in the depth ranking.
These two nodes have high centrality scores for betweenness and closeness (20) and degree (33).
The last Fig.~\ref{fig:deepest-karate} shows the node of the Zachary Karate Club network that has the top $\PTD$ score (i.e. the generalised median), according to diffusion distance embedding (and its parameters $t, p$) and shortest-path distance embedding with parameter $p$.

\subsubsection{2010 Network Scientists Network.}

The second social network analysed in this work is the network of Network Scientists from 2010 \citep{Edler2013}.
It is a co-authorship network, where 522 scholars are linked by weighted edges representing the relation $u \sim v$ if and only if there is a paper with $u, v$ among the authors.
The weight of edge $\{u, v\}$ depends on the number of papers co-authored by $u, v$.

\begin{figure}[th]
  \centering
  \includegraphics[width=.95\textwidth]{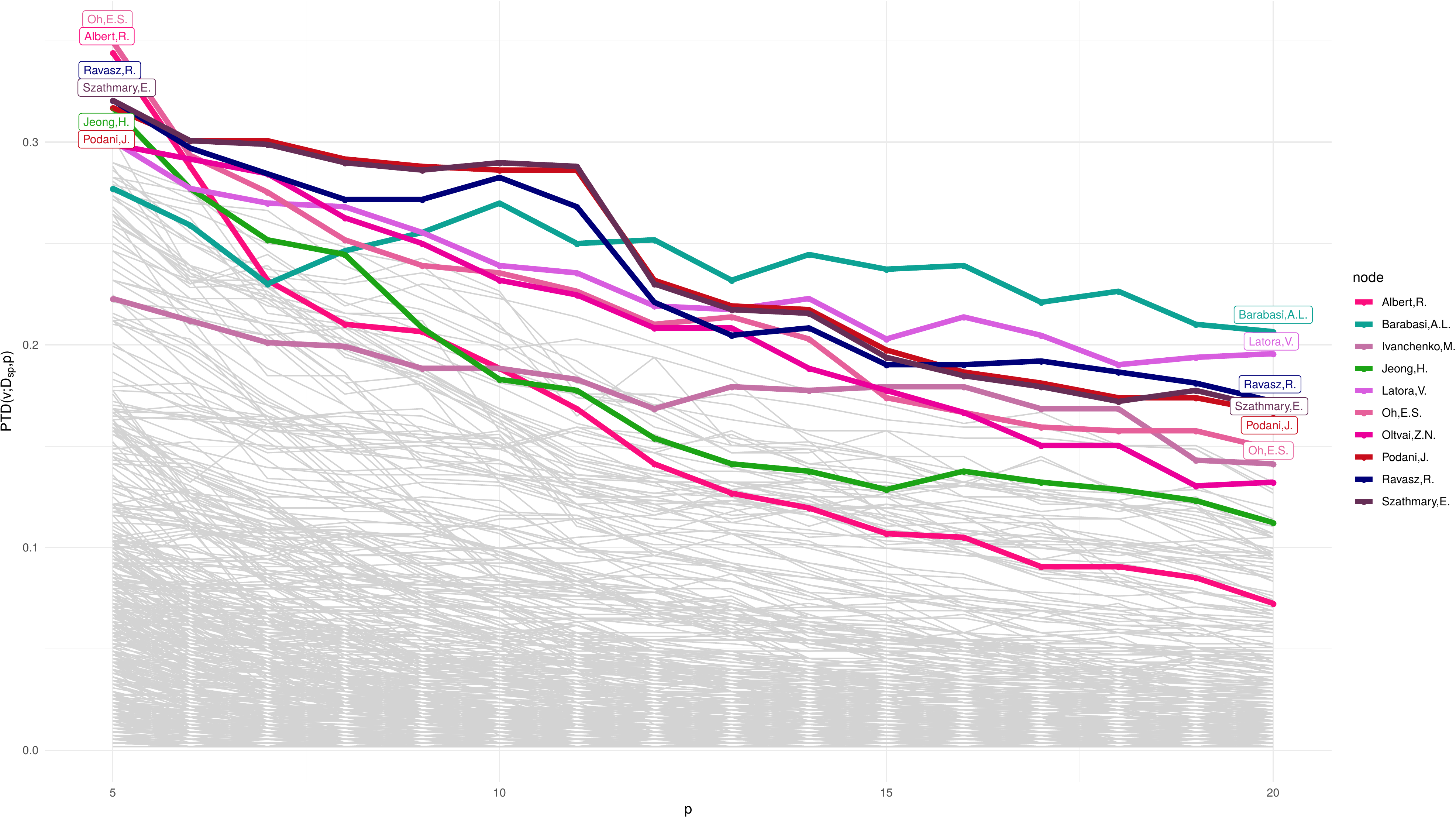}
  \caption{Network depth patterns for the collaboration network of Network Scientists 2010, according to $D_{sp}$ embeddings in $\R^p$, $p = 5, \dots, 20$. At each $p$, nodes in the depth region $R_{\alpha, n}$ of order $\alpha=0.99$ have a coloured line. The depth range decreases as $p$ grows, because the sample is more and more diluted in high dimensional spaces. Oh, E.S. and Albert, R. have high depth in $\mathbb{R}^p$ for small values of $p$ but the most persistent pattern has Barabasi, A.L. as median node.}
  \label{fig:netsci-sp}
\end{figure}

\begin{figure}[th]
  \centering
  \includegraphics[width=.95\textwidth]{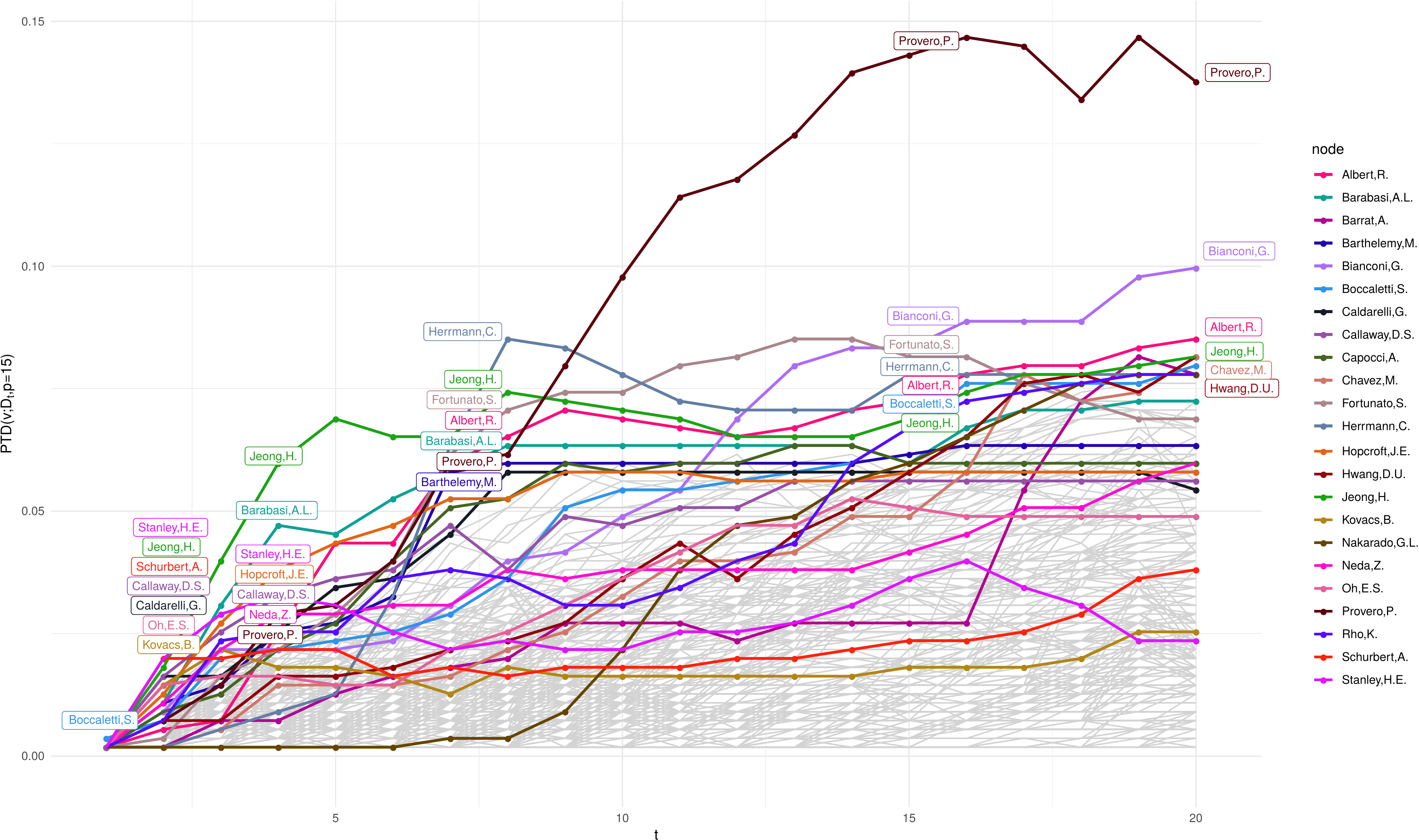}
  \caption{Depth patterns in $\R^{15}$ diffusion embeddings with $t = 1, \dots, 20$, for the collaboration network of Network Scientists 2010. Nodes that lie in $R_{\alpha=0.995}$ for some value of $t$ have a coloured line, while for each value of $t$ the nodes in the central region for that $t$, $R_{\alpha=0.995}(t)$, are also labelled. At small time scales the median node is, at first, Boccaletti, S., and then goes from Stanley, H.E. to Jeong, H.. After $t=10$ Provero, P. has the largest depth value.}
  \label{fig:netsci-Dt}
\end{figure}

Depth patterns in embeddings depending on $D_{sp}$ and $D_t$ (Fig.~\ref{fig:netsci-sp}-\ref{fig:netsci-Dt}) are, as expected, very different.
$\PTD(D_{sp}; p)$ summarises the information enclosed in the matrix $D_{sp}$, as closeness centrality.
As a matter of fact, if we look at the five nodes with top scores according to closeness -- Ravasz, R. Jeong, H. Podani, J. Szathmary, E. and Barabasi A.L. -- we already find a strong overlap.
The correlation analysis in subsection \ref{ssec:correl-analysis} highlights this pattern.

\begin{figure}[th]
\centering
\includegraphics[width=.8\textwidth]{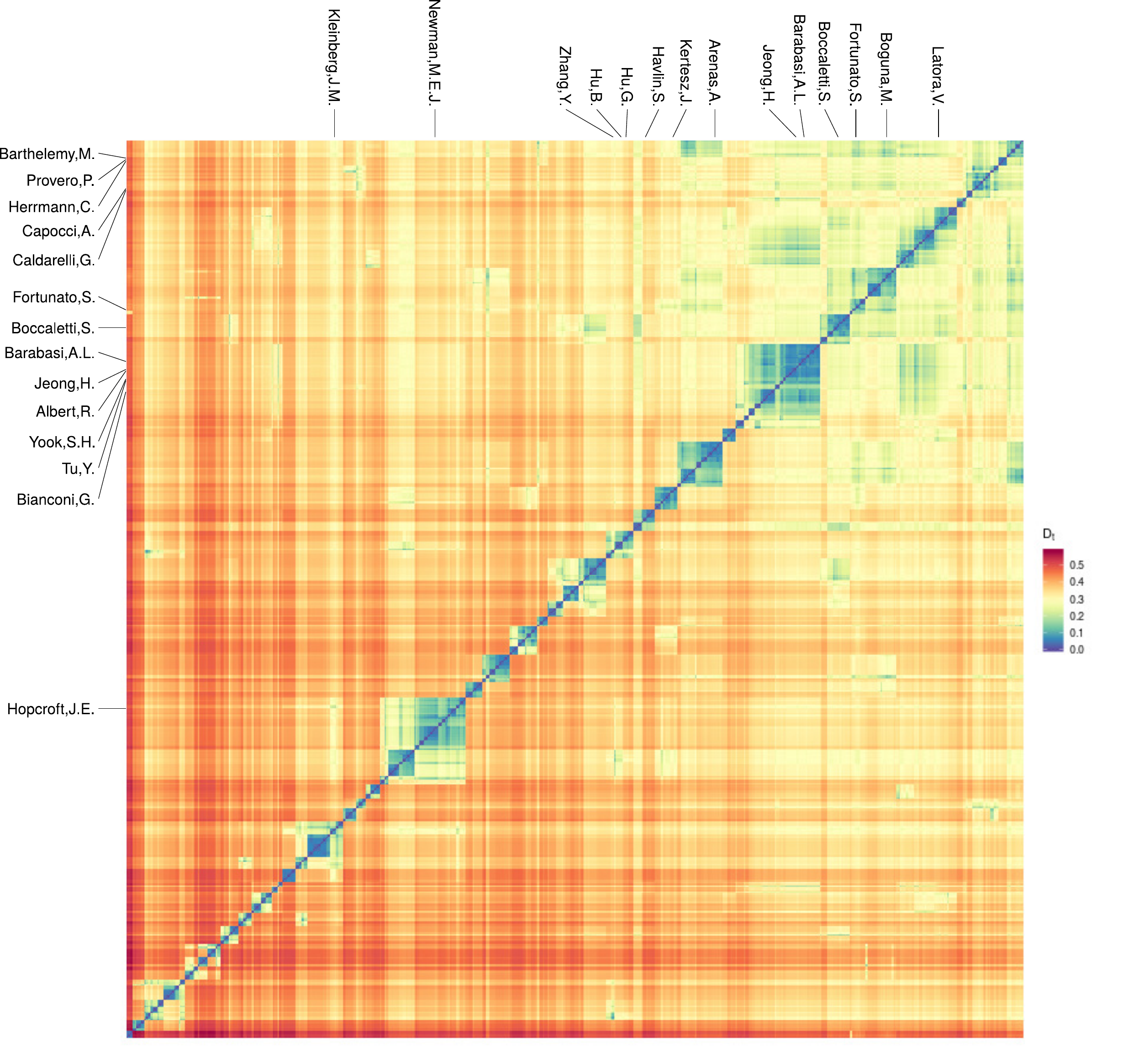}
\caption{Average diffusion distance matrix for the collaboration network of Network Scientists 2010. On the y-axis the 2.5\% deepest nodes (depth region of order 97.5-percentile), on the x-axis the top 2.5\% scholars w.r.t. betweenness centrality. We can see that there are several blue blocks across the diagonal representing groups of closely connected nodes. Deep nodes in the diffusion embedding, on the other hand, are not extremely close to anyone, but quite near to everyone.}
\label{fig:netsci-dist}
\end{figure}

On the other hand, the ranking based on $\PTD(D_t; t, p)$ may appear quite counterintuitive unless we have a clear picture of the information summarised by our network depth.
Fig.~\ref{fig:netsci-dist}, together with the four dimensions topology characterisation of our $\PTD(D_t; t, p)$, shades light on the results: the diffusion distance quantifies how difficult it is that two random walkers generated in two different nodes meet somewhere in the network.
In terms of this co-authorship network, a node which is sufficiently near to all others corresponds to a scholar which has a multifaceted co-authoring history.
Several blocks are clearly visible in the heatmap Fig.~\ref{fig:netsci-dist}, but remember that depths ignore multi-modality, exactly as the usual median.

\subsection{Biological Networks}

\subsubsection{Drosophila's Connectome}

The \textit{Drosophila} connectome \citep{Shih2015} is a directed weighted network with 49 nodes, 43 of which are local processing units (LPUs), while the remaining 6 are interconnecting units.
Edge weights represent the strength of total connections in terms of neuronal fibres between neurons inside LPUs.
\textit{LPUs} have their own local interneurons populations (LNs) whose fibres are limited to that region and exchange information with other LPUs via bundled tracts.
\textit{Interconnecting units} lack an LNs population and appear to relay information to other LPUs unmodified by any local network interactions.
In \citep{Shih2015} the authors show that this has a modular structure with five functional clusters: four of them correspond to sensory modalities and the fifth (called ``pre-motor centre'') is the integration centre for visuo-locomotor behaviours, which is important for decision-making. The inter-modular connectors are SPP, DMP, VLP-D, IDFB.

\begin{figure}[!h]
\begin{minipage}{.7\linewidth}
\centering
\subfloat[In-going geodesics]{\includegraphics[width=\textwidth]{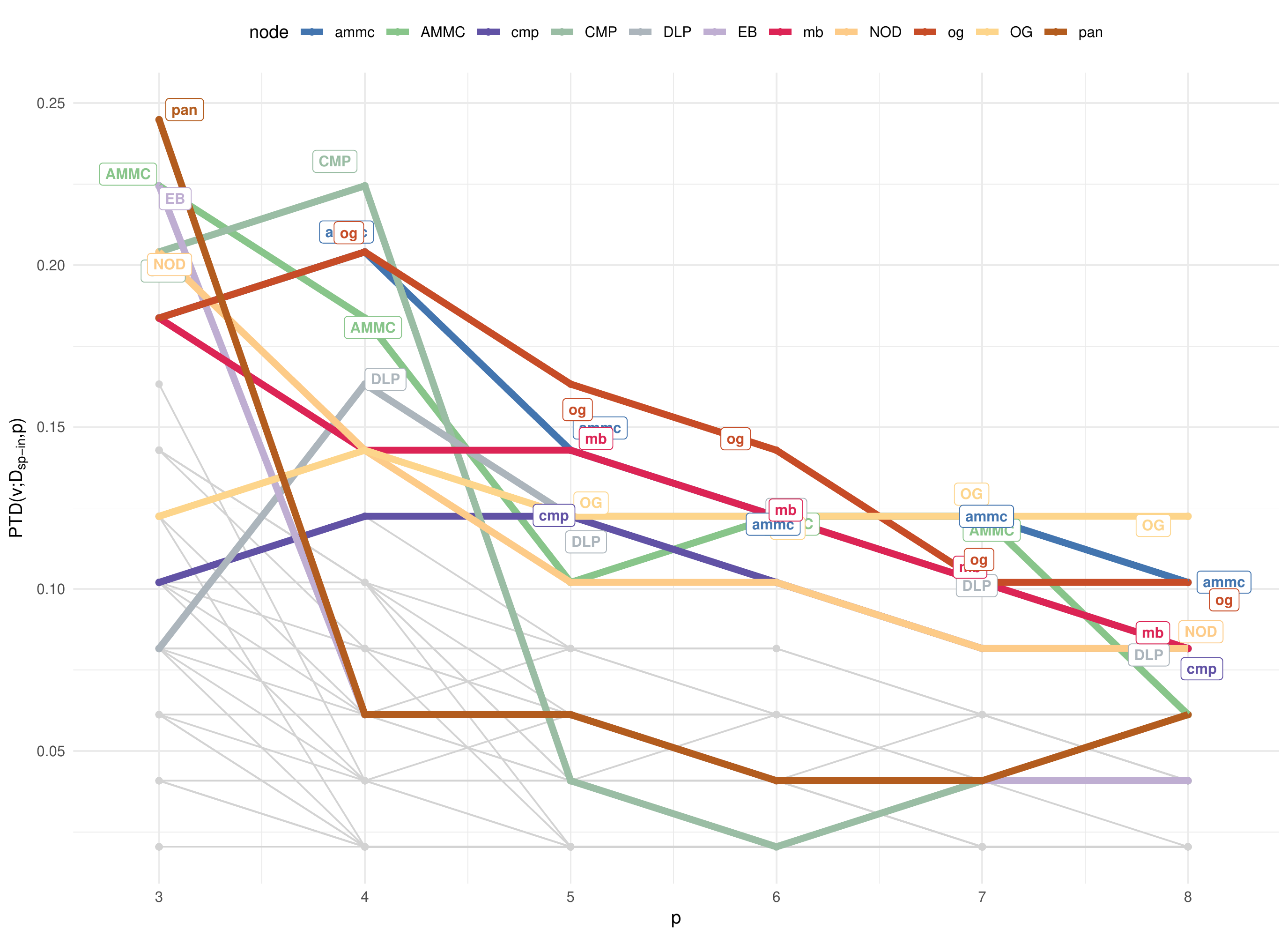}}
\hfill
\subfloat[Out-going geodesics]{\includegraphics[width=\textwidth]{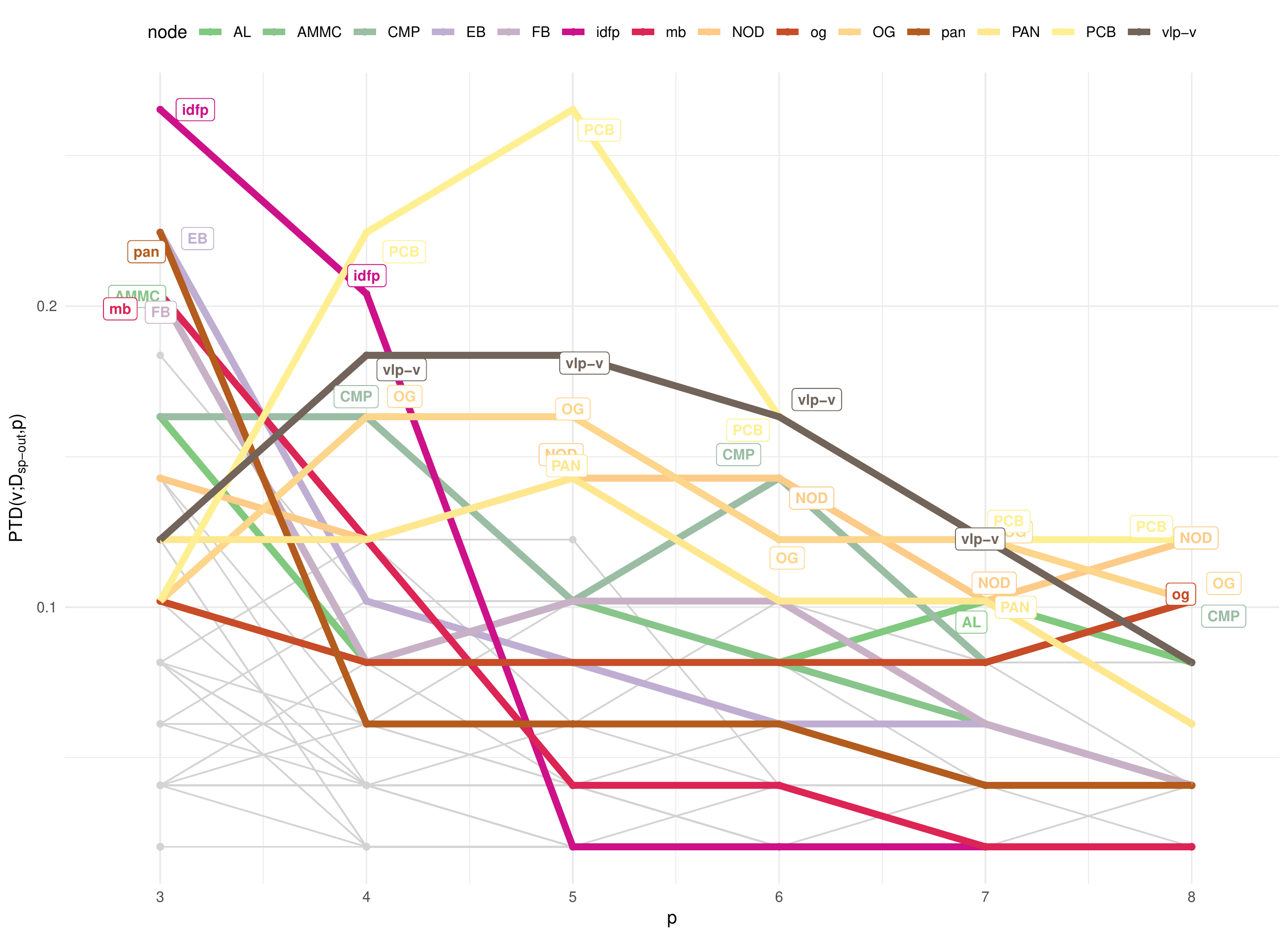}}
\end{minipage}%
\begin{minipage}{.26\linewidth}
\subfloat[\textit{Polarity}]{\label{sfig:polarity}\includegraphics[width=\textwidth]{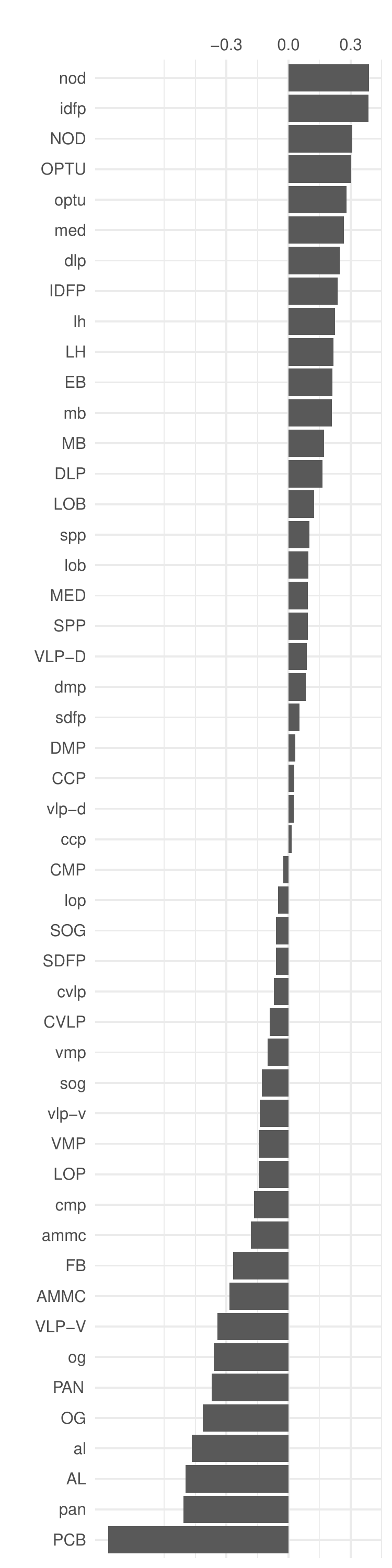}}
\end{minipage}%
\caption{$\PTD(D_{sp}, p)$ depth pattern w.r.t. directed shortest-paths for the Drosophila's connectome. LPUs in the $R_{\alpha=0.9, n}$ have coloured lines.}
\label{fig:dros-sp}
\end{figure}

In Fig.\ref{fig:dros-sp} we show the depth patterns in in-going and out-going geodesic embeddings with $p\geq 3$.
Since the directed shortest-path distance matrix is no more symmetric we get the space configuration of nodes through unfolding \citep{Cox2000}.
Lines carry a label if the corresponding node belongs to the depth region of order $\alpha=0.90$.
We can compare the centrality patterns with the ranking based on node polarity, the difference between in- and out-strength; positive and negative values indicate ``recievers'' and ``senders'' respectively. Like degree, this is a very local measure, instead the PTD depends on the global distance induced space configuration.
Observe that the aggregate $\PTD(D_{sp-out})$ depth trimmed region of order $\alpha=0.90$ contains the following LPUs: PCB, AL, CMP, NOD and vlp-v; while for $\PTD(D_{sp-in})$ the deepest LPUs are mb, og, NOD, AMMC, ammc.
PCB, protocerebral bridge, and noduli(NOD), belong to the integration center for visuo-locomotor behaviours in all anthropods; antennal lobes (AL) and antennal mechano-sensory and motor center (AMMC) (with their respective counterparts in the opposite hemisphere) are first-layer sensory (resp. for olfaction and hearing) LPUs, while OG, Optic Glomerulus, is one of six interconnecting units in the Drosophila's brain, which appear to relay information to other LPUs unmodified by any local network interactions \citep{Shih2015}.
It is thus not surprising that OG/og is among the deepest node both w.r.t. in-coming and out-going connectivity.
AMMC is a strong sender according to node polarity, but it also in the downstream of ammc and other senders (PAN/pan, AL/al, OG/og among others) and of LPUs with high degree/strength like DMP/dmp.

\begin{figure}[!h]
\centering
\subfloat[$\PTD(D_{t}; t, p=3)$ depth pattern w.r.t. diffusion distances for $t \in 1, \cdots, 20$.]{\label{fig:dros-Dt-p3}\includegraphics[width=0.8\textwidth]{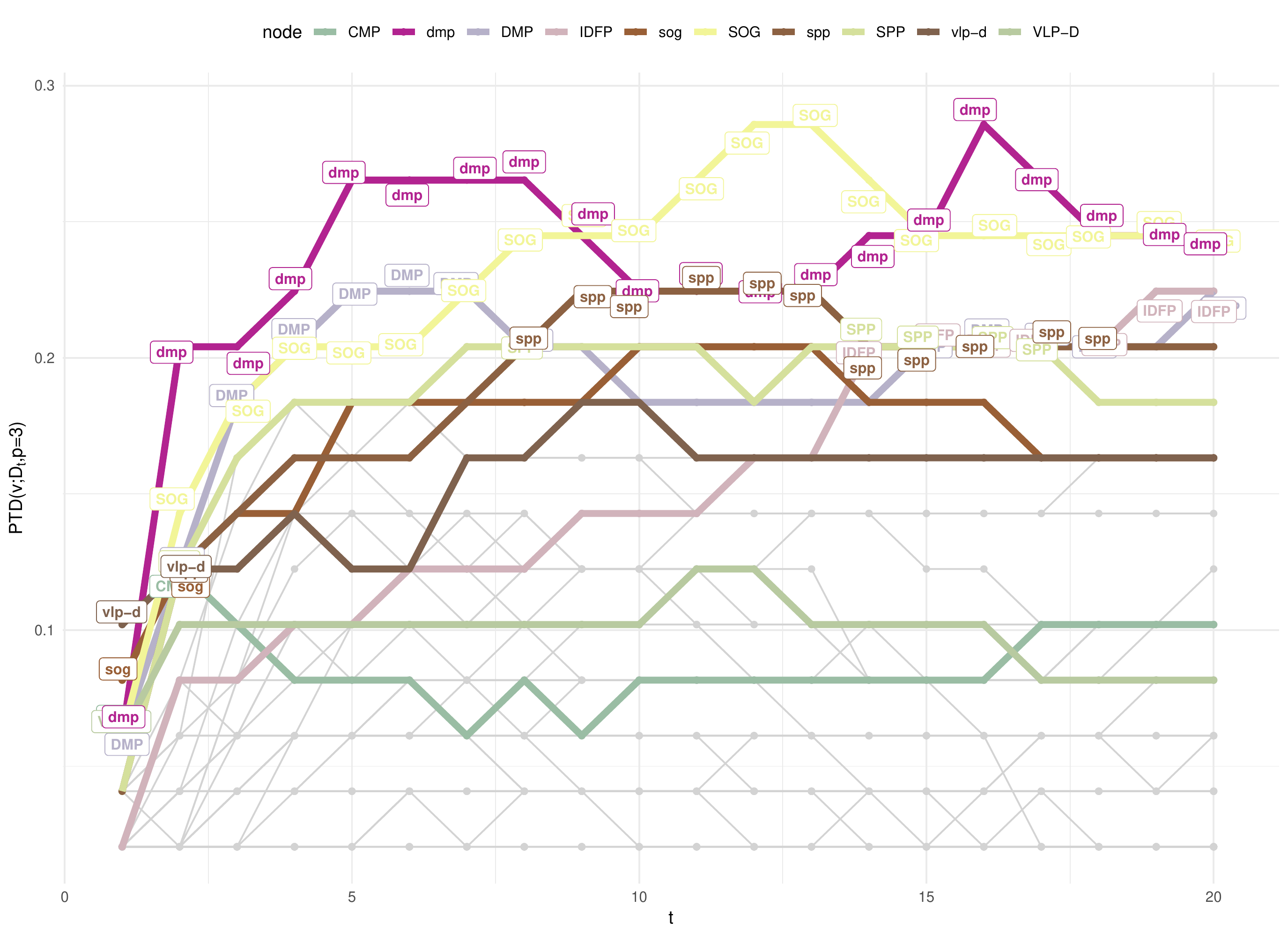}}
\subfloat[Global Centrality \citep{Shih2015}]{\label{fig:GC}\includegraphics[width=0.13\textwidth]{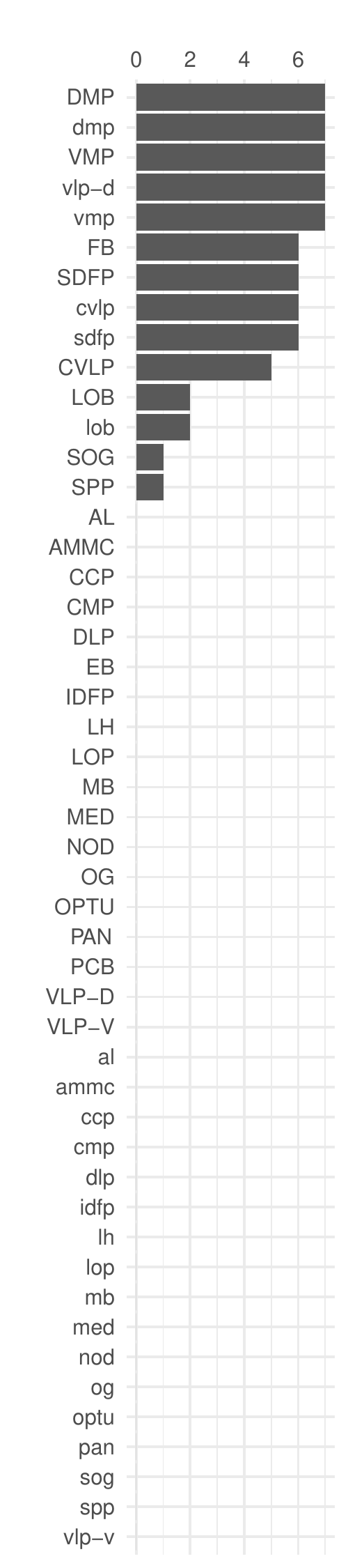}}
\caption{Network depth based on diffusion geometry and Shih's Global Centrality scores for the Drosophila's connectome.}
\end{figure}

A different scenario appears if we look at depth patterns w.r.t. diffusion embeddings in Fig.\ref{fig:dros-Dt-p3}: the top $10\%$ of the depth is quite stable with dmp (Dorsomedial Protocerebrum) being the deepest node for most of the time. If we average over the whole parameter space, so that we do not even need to decide which pair $(t, p)$ to use, and look at the aggregate $\PTD(D_t)$ the depth trimmed region of level $\alpha = 0.75$ consists of: DMP, FB (Fanshaped Body), IDFP (Inferior Dorsofrontal Protocerebrum), SOG (Subesophageal Ganglion), SPP (Superpenduncular Protocerebrum), VLP-D (Ventrolateral Protocerebrum, Dorsal part), ccp (Caudalcentral Protocerebrum), dmp, idfp, sog, spp, vlp-d, vmp (Ventromedial Protocerebrum).
Interestingly there is a large overlap with Shih's Global Centrality (GC), the result of a consensus among centralities: for each type of measure the top 12 out of 49 nodes (top quartile) earned 1 point and then by summing up one gets the node GC.
One significant methodological difference between aggregate $\PTD(D_t)$ and GC is that we do not arbitrarily bestow points, we use sample depth quantiles, getting a continuous measure that can be interpreted in the same way as all other depths.

\subsubsection{Protein-protein network}

Protein-protein interaction network of the yeast S. cervisiae \citep{Jeong2001}.
This is a medium size network (1870 nodes, from which 1458 in the largest connected component), edges are undirected and unweighted, while nodes carry metadata on the protein lethality.
According with lethality-centrality hypothesis, central nodes are important for cellular functioning. Nodes with highest degree (top 0.5\%) are YDL100C (Non-Lethal), YLR423C (Non-Lethal), STE50 (Non-Lethal), SNP1 (Lethal), TEM1 (Unknown), LSM8 (Lethal), SEC17 (Lethal), Nup1 (Unknown).

\begin{figure}[th]
  \centering
  \includegraphics[width=\textwidth]{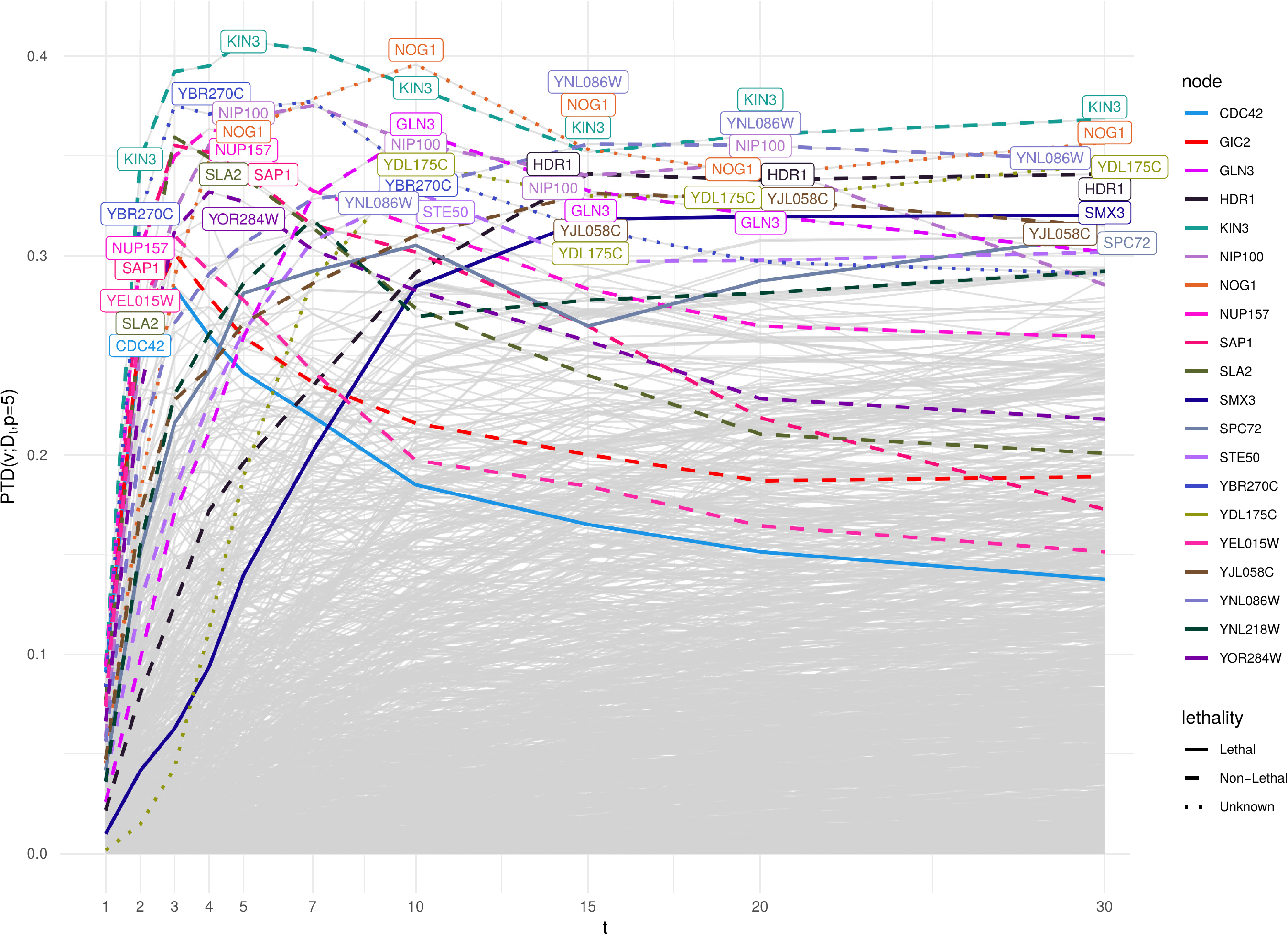}
  \caption{Depth patterns in $\R^{5}$ w.r.t. diffusion embeddings, for the protein-protein interaction network. Nodes that lie in $R_{\alpha=0.995}$ for some value of $t$ have a coloured line, while for each value of $t$ the nodes in the central region for that $t$, $R_{\alpha=0.995}(t)$, are also labelled. We can see that median and highly deep nodes are almost always Non-Lethal.}
  \label{fig:scervisiae}
\end{figure}

Fig.~\ref{fig:scervisiae} shows the depth patterns for $\PTD(D_t, t, p=5)$.
Without a detailed field knowledge it is meaningless to discuss on the importance of these proteins, however from the data we can infer that the protein-protein interaction network is disassortative, which means that Lethal protein are not necessarily connected only to other Lethal proteins. This implies that the spatial distribution of points embedded according to diffusion (or shortest-path distance) does not reflect the Lethality-communities.

\subsection{Infrastructure Networks}

The Euro Roads infrastructure network \citep{Subelj2011} is the international E-road network, located mostly in Europe.
It is an undirected network, with nodes representing cities and edges between nodes denoting E-road connections. We focus on its largest connected component, containing $N=1039$ (over the total 1177) vertices.

This real network lives in $\mathbb{R}^3$, but as we try to embed it into $\mathbb{R}^3$, solving analytically the classical MDS problem with distance matrix $D_{sp}$ and $p=3$, the result is poor with an explained variance of 74.9\%. This agrees with our expectations, since the shortest-path distance is not equivalent to the Euclidean distance. Using the majorisation approach described in \ref{ssec:net-emb} and the non-metric MDS, a more appropriate embedding dimension according to stress evaluation is $p=5$, which yields a stress-1 of 0.038. The cities in the depth region $R_{\alpha=0.99}$ (top 1\%) based on $\PTD(D_{sp}; p=5)$ are reported in Tab.~\ref{tab:euroad}, together with the top 1\% of the common centrality measures used in this study.

\begin{table}[th]
\centering
\scriptsize
\begin{tabular}{lllll}
$\PTD(D_{sp}; p=5)$ & Eigenvector & Closeness & Betweenness & Fragmentation  \\ \hline
Kingston upon Hull & Le Havre & Leeds & Kingston upon Hull & Urziceni \\
Korczowa & Arlon & Kingston upon Hull & Stara Zagora & Ishim \\
Osnabr\"{u}ck & Eupen & Esbjerg & Urziceni & Memmingen \\
Kumanovo& Tallinn    & Osnabr\"{u}ck    & Leuven  & Kingston upon Hull \\
Görlitz & Eindhoven  & Voss  & Brussels& Leeds \\
\r{A}ndalsnes    & Leuven   & Belgrade& Esbjerg & Irkeshtam    \\
Dej   & Paris    & Brussels& Leeds & Stara Zagora   \\
Voss  & Bastogne   & Panev\.e\v{z}ys    & Panev\.e\v{z}ys    & Guzar \\
Chemnitz& Sz\'ekesfeh\'erv\'ar & Rovaniemi    & Incukalns    & Reni  \\
Manchester   & Dover    & Stara Zagora   & Riga  & Leuven  \\
Brussels& Sion& Ruse  & Turku & Aarhus
\end{tabular}
\caption{Top 1\% of several centralities for the Euro-road network.}
\label{tab:euroad}
\end{table}

\begin{figure}[th]
  \centering
  \includegraphics[width=\textwidth]{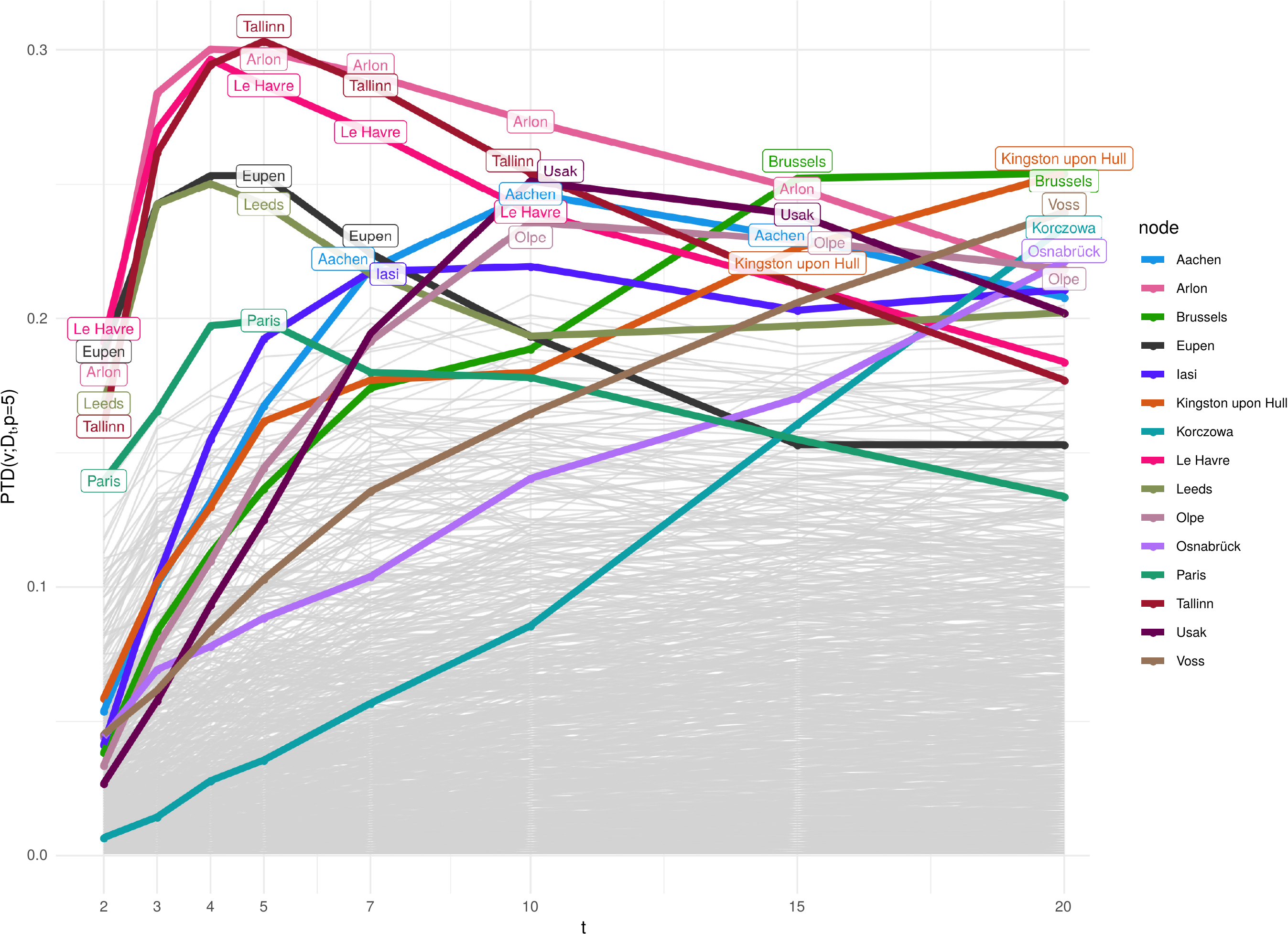}
  \caption{Depth patterns in $\R^{5}$ w.r.t. diffusion embeddings, for the road network. Nodes that lie in $R_{\alpha=0.995}$ for some value of $t$ have a coloured line, while for each value of $t$ the nodes in the central region for that $t$, $R_{\alpha=0.995}(t)$, are also labelled.}
  \label{fig:euroad-Dt}
\end{figure}

We also embed the Euro-road network based on diffusion distance, with $t \in \{1, \dots, 20\}$ and $p \{\in 2, \dots, 30\}$. Fig.~\ref{fig:euroad-Dt} shows the depth patterns for the network depth in $\mathbb{R}^5$ and a subset of diffusion times.
We can see a claer shift from ``local'' to ``global'' centrality of cities as $t$ grows. For small $t$ the nodes in the top 0.5\% are Le Havre, Eupen, Arlon, Leeds, Tallinn and Paris (this remains true also for other values of $p$), with a high overlap with the eigenvector centrality.
There is, however, a second stack of lines (nodes) with low depth values for small $t$, which gain depth as $t$ grows.
When $t=20$ the depth central region of level $\alpha=0.995$ looks very different: the median nodes are Kingston upon Hull and Brussels, followed by Voss, Korczowa, Osnabr\"{u}ck and Olpe. This ranking is more similar to the rankings given by the betweenness, closeness and even $\PTD(D_{sp}; p=5)$.

The similarity between $\PTD(D_{t}; t, p)$ and $\PTD(D_{sp}; p)$ is a consequence of the sparsity of the network. The diffusion distance between two nodes is small if they are connected by many or by short paths and since the network has a low edge density, the two metrics have similar patterns.

\subsection{Correlations Analysis of Centrality Measures}\label{ssec:correl-analysis}

In this subsection we perform a simple correlation analysis, by means of Spearman's rank correlation coefficients.
For three of the previously studied networks and for each pair $C_1, C_2$ of centrality measures, we compute the Spearman's rank correlation coefficient $r_s$; additionally a two-sided statistical test of level $\alpha = 0.05$ for $H_1: ~ r_s \neq 0$, see \citep[Sec.~8.5]{Hollander2013} for statistical details.

\begin{figure}[!h]
\begin{minipage}{.7\linewidth}
\centering
\subfloat[Drosophila's Connectome CMs Spearman Correlation Matrix]{\includegraphics[width=\textwidth]{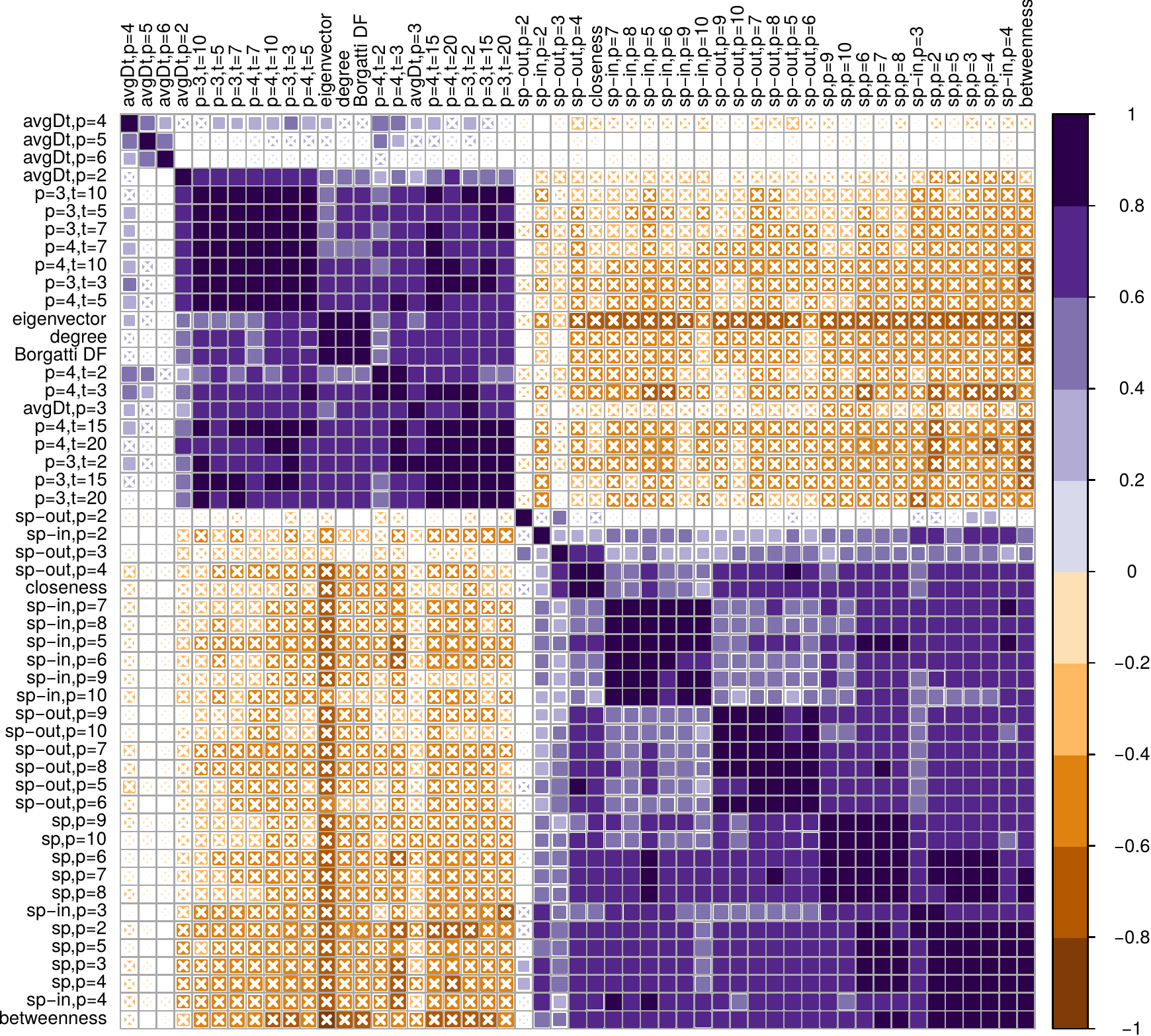}}
\end{minipage}%
\begin{minipage}{.3\linewidth}
\subfloat[Karate Club]{\includegraphics[width=\textwidth]{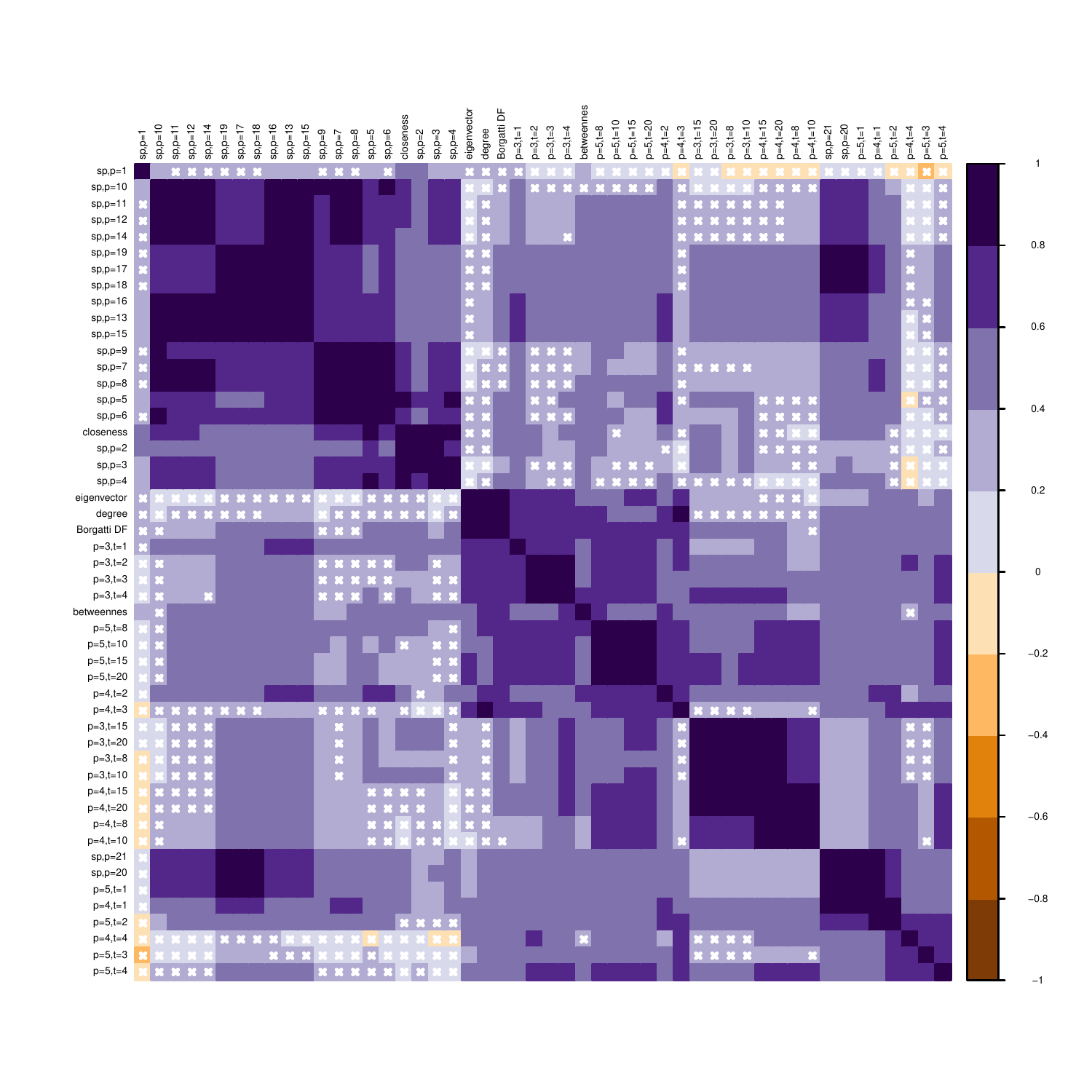}}\\
\subfloat[Network Scientists 2010]{\includegraphics[width=\textwidth]{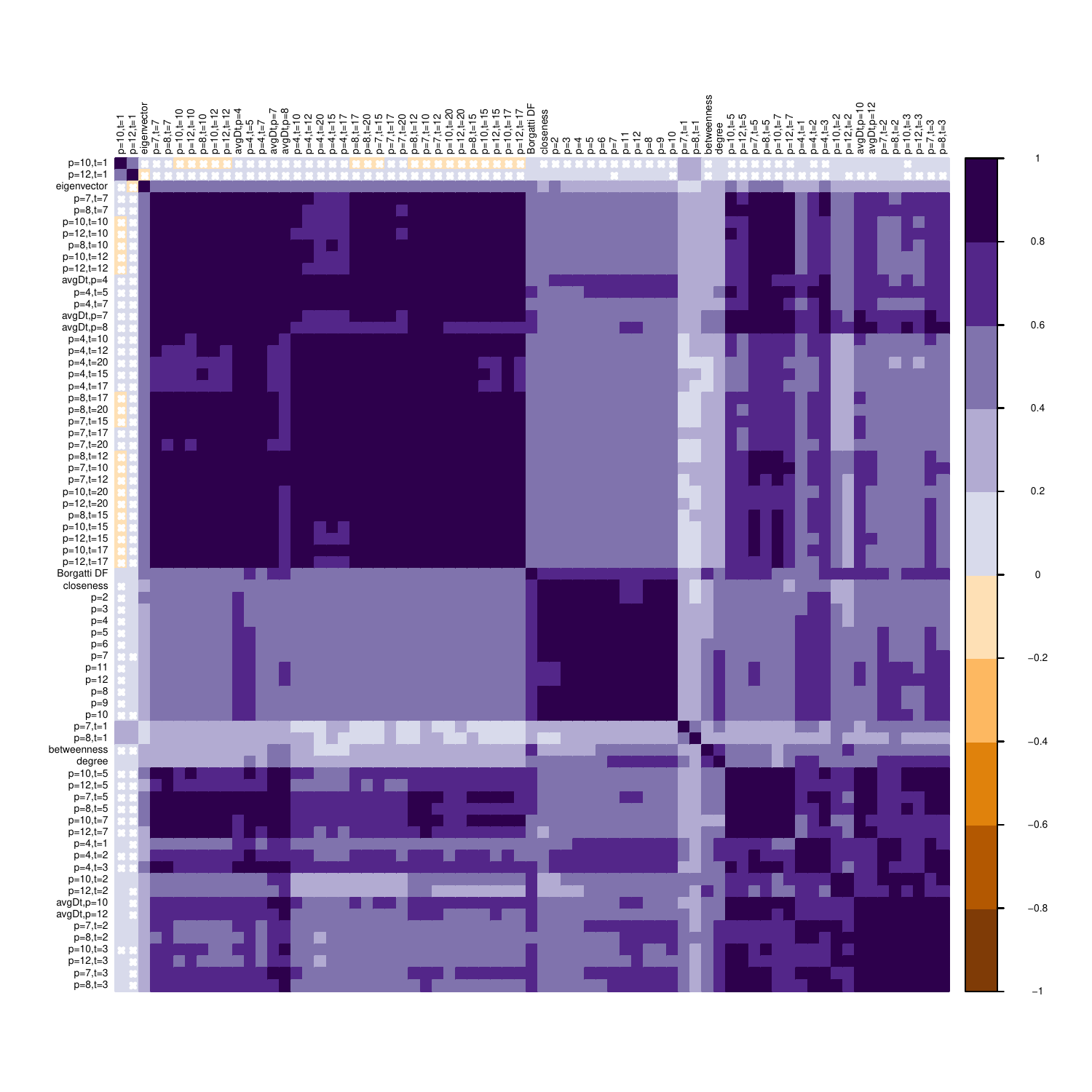}}
\end{minipage}%
\caption{Spearman's correlation among common centralities and network depths w.r.t. different parameters $D, p, t$. According to a statistical test of order $\alpha=0.05$, correlation coefficients that are not significantly different from 0 are crossed out. The common patterns observed in all three correlation matrices are: (i) a block structure with strong correlations among depths based on the same distance matrix and similar embedding dimensions, that is weaker correlations between $\PTD(D_{sp; p})$ and $\PTD(D_t; t, p)$; (ii) $r_s$ coefficient tends to be higher for measures falling in the same category according to Borgatti's framwork \citep{Borgatti2006}, e.g. closeness in strongly correlated to $\PTD(D_{sp}, p)$, while degree and eigenvector have generally higher correlations with depths in diffusion embeddings.}
\label{fig:cor-all}
\end{figure}

Fig.~\ref{fig:cor-all} shows the correlation matrices for the three networks analysed in this paper, which share a common pattern.
Indeed, blocks of strongly correlated measures emerge: for different values of parameters $p$ and $t$ the families $\{\PTD(D_{sp}; p)\}_p$ and $\{\PTD(D_t; t, p)\}_{t, p}$ have strong inter-family correlations and weak or even non-significant intra-families correlations.
Further, closeness falls generally in the group $\{\PTD(D_{sp}; p)\}_p$ while degree and eigenvector correlate more with $\{\PTD(D_t; t, p)\}_{t, p}$, as expected from the categorisation of centralities using the graph-theoretic framework of centrality measures \citep{Borgatti2006}.

\begin{figure}[th]
  \centering
  \includegraphics[width=\textwidth]{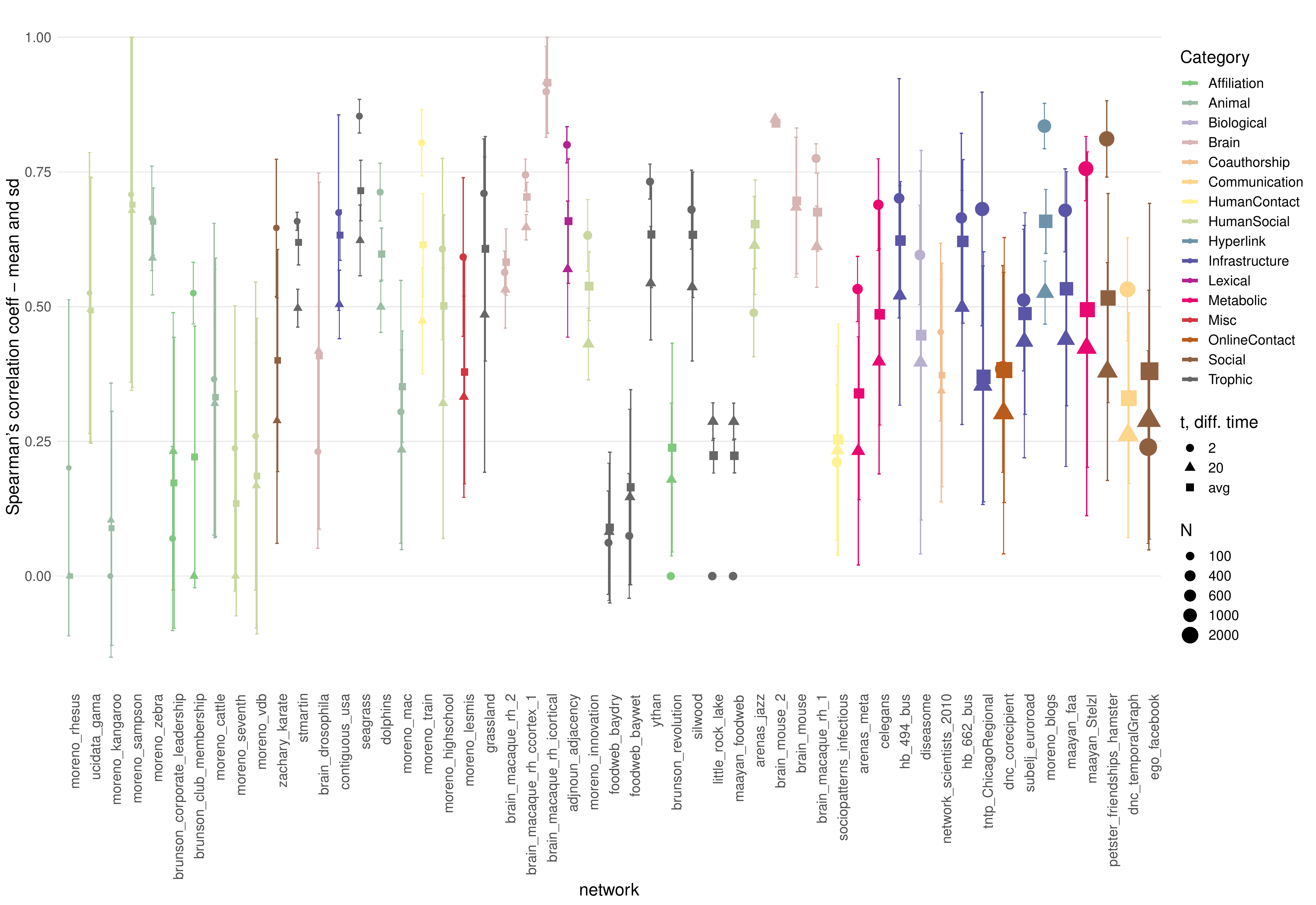}
  \caption{Average Spearman's correlation between the network depth and common centralities measures. More than 50 networks have been embedded according to the diffusion distance for $t=2$ (short walks, local information) and $t=20$ (local information is integrated resulting in a global similarity measure) and to the average diffusion distance, $\bar{D}_t$. The Projected Tuckey Depth of the resuting data clouds has been computed and their Spearman's correlation w.r.t. other common centralities (degree, eigenvector, PageRank, etc) has been computed. The results are here summarised, through the mean correlation and standard deviation (error bars).}
  \label{fig:corr-exten}
\end{figure}

Fig.~\ref{fig:corr-exten} shows an extensive comparison in terms of mean Spearman's correlation between the network depth in different diffusion embeddings and common centralities, on more than 50 real networks (mostly from KONECT \citep{KONECT} repository).
Correlations coefficients for which the $p-$value for $H_0: r_s = 0$ is larger than 0.05 have been set to zero.
The correlations do not show clear category- or size-based patterns. Correlations are generally moderate (below 0.7). One network stands out particularly: the interareal cortical network of the macaque \citep{Markov2013}.

\section{Conclusions}\label{sec:conclusions}

We have shown that it is possible to extend the definition of median to structured data like networks, through a newly defined statistical data depth.
This also provides a continuous centrality measure, the network depth, which depends on a network embedding step. This step introduce one parameter $p$ -- in case of an embedding based on geodesics -- or two parameters $p, t$ if we use the diffusion distance.
The dimension of the embedding space $p$ affects the quality of the network depth, but a reasonable $p$ can be selected without much effort, as we have shown at the end of Sec.~\ref{ssec:net-emb} and in the applications.
The diffusion time $t$ is more than a free-parameter, it is a resolution parameter, it allows us to move from the micro to the macro-scale, underlying locally central nodes, e.g. centres of communities, or community connectors (see Drosophila's connectome in Sec.~\ref{sec:applications}).

The network depth arises from an unprecedented approach: the definition of the centrality does not follow the definition of the \textit{importance} a node, it is the application of a pure statistical methodology, the generalisation of order relations through depth functions.
For this reason and to prove that the network depth is a network centrality measure, we place it in the graph-theoretical taxonomy of centralities by Borgatti and Everett, theoretically inferring its properties. We have shown that the $\PTD$ is a radial measure and hence it looks for a global centre.
Furthermore, which distance we choose to evaluate on the network has an important role in determining the distribution of points in space. Therefore, the depth-based node ranking has to be interpreted in light of the metric considered.

In this work we have analysed different social, biological and infrastructural systems to understand if the centrality score induced by the Projected Tukey depth reveals nodes with special or relevant meaning not detected by more traditional centrality measures, such as degree and eigenvector -- radial, volume measures --, closeness -- a radial, length centrality summarising information given by length of shortest paths --, or betweenness and Borgatti's fragmentation, two medial measures of centrality focusing on paths passing through each node.
When diffusion geometry is used to embed the networks, we find our depth to unveil central nodes with a heterogeneous connectivity across the network. In other words, nodes that, instead of having a strong belonging to a community, are well-mixed inside the network structure.

Our results provide a grounded framework for the definition of median node and contours in complex networks.

\bibliographystyle{comnet}

\end{document}